\documentclass[aps, floatfix, a4paper, nofootinbib,
superscriptaddress, 11pt]{revtex4}

\usepackage[english]{babel} 
\usepackage{graphicx,float,tikz}
\usepackage[all]{xy}
\usepackage{amsmath,upgreek}
\usepackage{amsfonts}
\usepackage{amsthm}   
\usepackage{bbold}
\usepackage{bm}
\usepackage{amssymb}
\usepackage{color}
\usepackage{epsfig}		
\usepackage{graphicx,epstopdf}
\usepackage{subfigure}
\usepackage{pdfpages}
\usepackage{dsfont}
\usepackage{float}
\usepackage{hyperref}
\usepackage{nameref}
\usepackage{bookmark}
\usepackage{xcolor}

\usepackage{multirow}
\definecolor{green1}{RGB}{0,128,0} 
\hypersetup{backref=true,pagebackref=true}
\hypersetup{%
  colorlinks = true,
  linkcolor  = blue,
  citecolor = blue,
}
\usepackage{bookmark,textgreek}
\hypersetup{hidelinks,hyperindex=true,colorlinks=true,breaklinks=true,urlcolor=blue}
\hypersetup{%
  colorlinks = true,
  linkcolor  = blue
}

\begin{document}

\title{Dispersion Relations and Pole-Skipping in a Holographic Charmonium Model with Rotating Plasma}

\author{Luiz  F. Ferreira}
\email{lffaulhaber@gmail.com}
\affiliation{Instituto de Física y Astronomía, Universidad de Valparaíso, A. Gran Bretana 1111, Valparaíso, Chile}

\begin{abstract}

 In this paper, we employ a bottom-up holographic QCD model to investigate the dissociation of charmonium states moving in a rotating medium by calculating their quasinormal modes. We begin by reviewing the holographic quarkonium spectrum at zero temperature. Then, we derive the equations of motion for heavy vector mesons propagating in a rotating plasma, separating the analysis into longitudinal and transverse directions relative to the wave vector. Additionally, we compute the charge diffusion constant in the rotating background and analyze the pole-skipping phenomenon, which emerges in the retarded Green’s function. Finally, we investigate the impact of rotation on the spin alignment of the $J/\psi$ state in the helicity frame, utilizing the spectral function obtained from the holographic framework.

 \end{abstract}

%\pacs{89.70.Cf, 11.25.Tq, 12.39.Mk}

\maketitle

\newpage
\medbreak

\section{Introduction}

One of the key probes for investigating quark-gluon plasma (QGP) produced in heavy-ion collisions is provided by quarkonium states. Their significance lies in the fact that the presence of a thermal medium leads to the partial dissociation of these hadronic states.  Beyond thermal effects, non-central collisions also generate a strong but short-lived magnetic field, as well as substantial orbital angular momentum on the order of $10^4$–$10^5\hbar$~\cite{Becattini:2007sr}. Although the magnetic field rapidly decays, the rotation persists along the axis perpendicular to the collision plane \cite{Jiang:2016woz}. This rotational effect significantly influences key aspects of QGP dynamics, including the phase diagram, dilepton production rates, and spin polarization. Consequently, understanding how rotation affects the dissociation of heavy mesons is essential to provide a comprehensive description of the QGP properties.

In recent years, gauge/gravity duality has emerged as an essential framework for exploring
diverse properties of heavy quarkonium systems. These include the spin alignment of quarkonium states \cite{Zhao:2024ipr, Ahmed:2025bwi}, quarkonium spectra \cite{Hohler:2013vca,Braga:2016wkm, Braga:2017bml, Braga:2018zlu, Braga:2018hjt, MartinContreras:2021bis, Zhao:2021ogc, Mamani:2022qnf, Zhao:2023pne, Toniato:2025gts}, thermal widths \cite{Finazzo:2013rqy, Braga:2016wzx, Zhao:2019tjq, Feng:2020qee}, and configuration entropy \cite{Braga:2017fsb, Braga:2018fyc, Braga:2020myi, Braga:2020hhs, Zhao:2023yry, Braga:2023fsc, MartinContreras:2023eft}. Additional relevant aspects have also been extensively explored in this framework, as discussed in Refs.~\cite{Chen:2023, Zhou:2020, Zhang:2023, Tahery:2023, Zhang:2011, Chu:2009, Hou:2008, Hou:2021, Chen:2018, Chen:2022, Zhu:2020, Wu:2022, Li:2021, Zhu:2023, Sun:2024, Wei:2023,Ferreira:2025lma}. 

In particular, the thermal behavior of heavy vector mesons can be studied by computing their quasinormal modes in the bottom-up holographic model proposed in Refs.~\cite{Braga:2018hjt, Braga:2018zlu}. Quasinormal modes correspond to the poles of retarded Green’s functions, and the associated gravitational perturbations provide dual descriptions of quasi-particle excitations in the boundary gauge theory. In this context, quasinormal modes describe the thermal behavior of vector mesons, which manifest as quasi-particle states in the dual field theory. The real part of the quasinormal  frequency is associated with the effective mass of the mesonic state, while the imaginary part determines its inverse lifetime, or equivalently, its thermal decay width.

In this study, we investigate how the rotation of the QGP influences the thermal spectrum of charmonium quasi-states at nonzero momentum. We begin by decomposing the analysis into longitudinal and transverse directions relative to the wave vector. We then analyze the spectral function for various values of momentum and angular velocity. Subsequently, we examine the dependence of the real and imaginary parts of the quasinormal frequencies on angular velocity at fixed momentum for both longitudinal and transverse cases. We also consider the complementary scenario in which the linear momentum is varied while the angular velocity is held fixed. In addition, we derive the diffusion constant in the presence of rotational effects and investigate the pole-skipping phenomenon in the holographic correlators of the charmonium model. Finally, we study the impact of rotation on the spin alignment of the $J/\psi$ state in the helicity frame, using the spectral function derived within the holographic framework.

The paper is organized as follows. In Sec. II, we revisit the holographic model for heavy vector mesons and derive the corresponding equations of motion. In Sec. III, we derive and study the spectral function in the rotating case. In Sec. IV, we compute the spectrum of quasinormal modes and evaluate the charge diffusion constant. Sec. V is devoted to the analysis of the pole-skipping phenomenon. In Sec. VI, we analyze the spin alignment. Finally, in Sec. VII, we present our conclusions and discussion.

\section{Holographic model for heavy vector mesons }

\subsection{Holographic model for heavy vector mesons at zero temperature. }

The bottom-up holographic model for heavy vector mesons considered here was originally developed in \cite{Braga:2017bml,Braga:2018zlu}. This model offers a significant and straightforward approach to describing the phenomenology of the charmonium state using holography. At zero temperature, the background geometry corresponds to the standard five-dimensional anti-de Sitter $(AdS_5)$ space-time:
\begin{equation}
 ds^2 \,\,= \,\, \frac{R^2}{z^2 }(-dt^2 + d\vec{x}\cdot d\vec{x} + dz^2)\,.
 \label{AdSmetric}
\end{equation}
On the gravity side, the heavy vector mesons are described by the fields $V_m = (V_\mu,V_z)\,$ ($\mu = 0,1,2,3$), whereas in the gauge theory framework, they are represented by the current $ J^\mu = \bar{\psi}\gamma^\mu \psi \,$. The action governing their dynamics is given by:
\begin{equation}
I \,=\, \int d^4x dz \, \sqrt{-g} \,\, e^{- \phi (z)  } \, \left\{  - \frac{1}{4 g_5^2} F_{mn} F^{mn}
\,  \right\} \,\,, 
\label{vectorfieldaction}
\end{equation}
where $F_{mn} = \partial_m V_n - \partial_n V_m$. The energy parameters are introduced through the background scalar field $\phi(z)$ \footnote{We emphasize that introducing the dilaton field is essential to break conformal invariance, as the original AdS/CFT correspondence describes a conformal theory on the field theory side.}.
The AdS/QCD setup, inspired by the gauge/gravity duality, considers a similar type of duality, where the conformal invariance is broken after introducing an energy parameter in the AdS bulk.

The scalar field proposed in Ref. \cite{Braga:2017bml} takes the following form
\begin{equation}
\phi(z)=c^2z^2+Mz+\tanh\left(\frac{1}{Mz}-\frac{c}{ \sqrt{\Gamma}}\right)\,.
\label{dilatonModi}
\end{equation}
The parameters $c$ and $\Gamma $ correspond, respectively, to the heavy quark mass and the string tension of the quark-antiquark interaction. In contrast, the interpretation of the third parameter, $M$, is more subtle.  Heavy vector mesons can decay via non-hadronic processes, where the final state consists of light leptons. These transitions, not mediated by the strong interaction, typically involve a substantial change in mass. The parameter $M$ effectively represents the mass scale relevant to such processes and is characterized by a matrix element of the form
\begin{equation}
 \langle 0 \vert \, J_\mu (0)  \,  \vert n \rangle = \epsilon_\mu f_n m_n \,, 
\end{equation}
which describes a transition from a meson in the $n$-th radial excitation state to the hadronic vacuum. Here, the decay constant $f_n$ is essentially proportional to this matrix element, and the large mass parameter $M$ allows for accurate fitting of the corresponding spectra.

Choosing the gauge \( V_z = 0 \), the equation of motion for the transverse (1,2,3) components of the field, denoted generically as \( V \), in momentum space reads:
\begin{equation}
\partial_z \left[ e^{-B(z)} \partial_z V \right] - p^2 e^{-B(z)} V = 0,
\label{eqmotion}
\end{equation}
where \( B(z) \) is  
\begin{equation}
B(z) = \log\left( \frac{z}{R} \right) + \phi(z)\,.
\label{B}
\end{equation}
The equation of motion (\ref{eqmotion}) admits a discrete spectrum of normalizable solutions, \( V(p,z) = \Psi_n(z) \), that satisfy the boundary condition \( \Psi_n(z = 0) = 0 \) for \( p^2 = -m_n^2 \), where \( m_n \) are the masses of the corresponding meson states. The eigenfunctions \( \Psi_n(z) \) are normalized according to:
\begin{equation}
\int^{\infty}_{0} dz \, e^{-B(z)} \Psi_n(z) \Psi_m(z) = \delta_{mn} \,.
\label{Normalization}
\end{equation}

The decay constants are proportional to the transition matrix element between the \( n \)-th excited vector meson state and the vacuum:
\begin{equation}
\langle 0 \vert J_\mu(0) \vert n \rangle = \epsilon_\mu f_n m_n \,.
\end{equation} These constants are calculated holographically, following the same procedure as in the soft wall model \cite{Karch:2006pv}:
\begin{equation}
f_n = \frac{1}{g_5 m_n} \lim_{z \to 0} \left( e^{-B(z)} \Psi_n(z) \right) \,.
\label{decayconstant}
\end{equation}

 The values of the parameters of the dilaton profile are chosen to provide the best fit for the masses and decay constants of charmonium, in comparison with the experimental data reported by the Particle Data Group~\cite{ParticleDataGroup:2014cgo}. The parameter values used to describe charmonium are: \begin{equation} c = 1.2\, \text{GeV}, \quad \sqrt{\Gamma_c} = 0.55\, \text{GeV}, \quad M_c = 2.2\, \text{GeV}. \label{parameters1} \end{equation} 
 
Table~\ref{Tab13} displays the masses and decay constants of charmonium states obtained from the model, compared with the corresponding experimental data. As can be seen, the holographic model employed here achieves an excellent fit for the decay constants relative to the experimental values. This agreement was the main motivation for developing this alternative holographic description. Conventional holographic approaches  fail to reproduce the experimentally observed behavior of decay constants. For instance, the hard-wall model proposed in Refs. \cite{Polchinski:2001tt,Boschi-Filho:2002wdj,Boschi-Filho:2002xih} predicts decay constants that increase with the radial excitation number, while the soft-wall model \cite{Karch:2006pv} yields degenerate values for all excitations.

 \begin{table}[h]
\centering
\begin{tabular}[c]{|c||c||c|}
\hline 
\multicolumn{3}{|c|}{  Holographic (and experimental)  Results for Charmonium   } \\
\hline
 State &  Mass (MeV)     &   Decay constants (MeV) \\
\hline
$\,\,\,\, 1S \,\,\,\,$ & $ 2943 \,\, (3096.916\pm 0.011)  $  & $ 399 \, (416 \pm 5.3)$ \\
\hline
$\,\,\,\, 2S \,\,\,\,$ & $  3959 \,\, (3686.109 \pm 0.012) $   & $ 255  \, (296.1 \pm 2.5)$  \\
\hline 
$\,\,\,\,3S \,\,\,\,$ & $  4757 \,\, (4039 \pm 1 ) $   & $198 \, ( 187.1  \pm 7.6) $ \\ 
\hline
$ \,\,\,\, 4S  \,\,\,\,$ & $ 5426\,\,  (4421 \pm 4)  $  & $ 169 \,  (160.8  \pm 9.7)$ \\
\hline
\end{tabular}   
\caption{Holographic masses and the corresponding decay constants for the Charmonium S-wave resonances. Experimental values inside parenthesis for comparison.}\label{Tab13}
\end{table}

Now, one might ask: why is it important to achieve an accurate fit for the zero-temperature decay constants when the goal is to provide a suitable description of finite-temperature thermal behavior? The answer becomes clear when we recall the connection between decay constants and the spectral function.

The spectral function at finite temperature corresponds to the imaginary part of the retarded Green’s function. At zero temperature, the two-point Green’s function can be written as a spectral sum over the masses $m_n$ and decay constants $f_n$ of the physical states:
\begin{equation}
\Pi (p^2)  = \sum_{n=1}^\infty \, \frac{f_n^ 2}{(- p^ 2) - m_n^ 2 + i \epsilon} \,.
\label{2point}
\end{equation}

The imaginary part of Eq.~(\ref{2point}) is expressed as a sum over delta functions, with coefficients given by the squared decay constants: $f_n^2 , \delta(-p^2 - m_n^2)$. As the temperature $T$ increases, these discrete peaks broaden in the spectral function, reflecting the reduced stability of the quasi-particle states. Therefore, ensuring that the model reproduces the correct decay constants at zero temperature is essential for any consistent extension to finite-temperature phenomena.

In Ref.\cite{PHENIX:2014tbe}, Figure 1 presents a compilation of results for quarkonium dissociation temperatures obtained from various theoretical approaches. For charmonium at zero chemical potential, dissociation occurs for the ratio $T/T_c$ in the range of 1.5 to 3.0. In the holographic model employed here (see Refs.\cite{Braga:2017bml,Braga:2018zlu}), the dissociation temperature of charmonium is found to be $1.7 \, T/T_c$, in good agreement with other theoretical predictions. In contrast, the soft-wall model predicts a dissociation temperature below the critical temperature, indicating a significant discrepancy with established results.

\subsection{Holographic model for heavy vector mesons in rotating plasma }

The description of rotating plasmas via holography can be approached in several ways. For example, Refs.~\cite{Arefeva:2020jvo,Golubtsova:2022ldm} model rotating quark--gluon plasma using the Kerr--\( AdS_5 \) metric, which describes a rotating black hole geometry. Alternatively, Ref.~\cite{Chen:2023yug} adopts an approach involving a boundary coordinate transformation in the large black hole limit of the Schwarzschild \( AdS_5 \) metric. 

In this work, we follow the formulation presented in Refs.~\cite{Chen:2020ath,Zhou:2021sdy,Braga:2022yfe}, where rotation is introduced via a Lorentz-like transformation applied to a cylindrical static metric.

The metric of \( AdS_5 \)/Schwarzschild in cylindrical coordinates is expressed as follows:
\begin{equation}
ds^2 = \frac{R^2}{z^2} \left( -f(z)\, dt^2 + (dx^1)^2 + (dx^2)^2 + (dx^3)^2 + \frac{dz^2}{f(z)} \right),
\label{metric2}
\end{equation}
with \( f(z) = 1 - \frac{z^4}{z_h^4} \), where \( z_h \) is the horizon position. By applying the Lorentz-like transformation presented in Refs.~\cite{BravoGaete:2017dso, Erices:2017izj},
\begin{eqnarray}
t &\rightarrow& \frac{1}{\sqrt{1 - l^2 \Omega^2}} \left(t - l^2 \Omega \varphi \right), \label{T1}\\
\varphi &\rightarrow& \frac{1}{\sqrt{1 - l^2 \Omega^2}} \left(\varphi - \Omega t \right), \label{T2}
\end{eqnarray}
where \( l \) is the radius of the hyper-cylinder and \( \Omega \) is the angular velocity of the rotating cylindrical black hole, the metric takes the form:
\begin{equation}\label{RotatingBHmetric}
ds^2 = \frac{R^2}{z^2} \left(-\gamma^2 f(dt - \bar{\Omega} l d\varphi)^2 + \gamma^2 (l d\varphi - \bar{\Omega} dt)^2 + \frac{dz^2}{f} + dx^2 + dy^2 \right),
\end{equation}
where \( \bar{\Omega} = \Omega l \) and the Lorentz factor \( \gamma \) is defined as:
\begin{equation}
\gamma = \frac{1}{\sqrt{1 - \bar{\Omega}^2}}.
\end{equation}

As a consequence of the coordinate transformation, this metric acquires non-zero angular momentum, whose expression was computed in Refs.~\cite{BravoGaete:2017dso, Erices:2017izj,Braga:2022yfe}:
\begin{equation}
L = \frac{2R^3}{k^2} \frac{\bar{\Omega}}{z_h^4 (1 - \bar{\Omega}^2)}\,.
\end{equation}
Thus, the metric~(\ref{RotatingBHmetric}) can naturally be interpreted as dual to a cylindrical slice of rotating plasma. Additionally, the temperature of the rotating \( AdS_5 \) background becomes~\cite{BravoGaete:2017dso, Erices:2017izj}:
\begin{equation} 
T = \frac{1}{\pi z_h} \sqrt{1 - \bar{\Omega}^2}\,.
\label{temp}
\end{equation}

To derive the equations of motion in terms of longitudinal and transverse electric fields, we introduce the coordinate system \( x^{\mu} = (x^0, x^1, x^2, x^3) = (t, \varphi, x^2, x^3) \), and consider the four-velocity at the boundary given by \( u^{\mu} = \gamma(1, \bar{\Omega}, 0, 0) \). With this setup, the metric takes the form:
\begin{equation}\label{metricrot2}
ds^2 = \frac{R^2}{z^2} \left( \left( -f(z)\, u_{\mu} u_{\nu} + h_{\mu \nu} \right) dx^{\mu} dx^{\nu} + \frac{dz^2}{f(z)} \right),
\end{equation}
where \( h_{\mu \nu} \) is the projector orthogonal to \( u_{\mu} \),
\begin{equation}
h_{\mu \nu} = \eta_{\mu \nu} + u_{\mu} u_{\nu}\,,
\end{equation}
and \( \eta_{\mu \nu} = (-1, 1, 1, 1) \) is the Minkowski metric at the boundary. Note that all boundary indices are raised and lowered with \( \eta_{\mu \nu} \).

The equations of motion for the heavy vector mesons are obtained by varying the action~(\ref{vectorfieldaction}) with respect to the metric~(\ref{metricrot2}). We adopt a plane wave ansatz:
\begin{equation}
V_{\mu}(z, x_\mu) = e^{-x_{\mu} \cdot k^\mu} V_\mu(z)\,,
\end{equation}
where \( k^\mu \) is the wave vector at the boundary, and we choose the radial gauge \( V_z = 0 \). The equations of motion then take the form:
\begin{eqnarray}
&& \partial_z \left( \frac{f}{z} \partial_z V_{\perp}^{\mu} - \frac{u^{\mu}}{z} \partial_z V_{\parallel} \right) + \frac{1}{f z} \left( k_{\parallel} \left( k_{\parallel} V^{\mu}_{\perp} - k^{\mu} V_{\parallel} \right) - u^{\mu} k_{\perp}^{\nu} \left( k_{\parallel} V_{\nu} - k_{\nu} V_{\parallel} \right) \right) \nonumber\\
&& \quad - \frac{k_{\perp}^{\nu}}{f z} \left( k_{\nu} V_{\perp}^{\mu} - k_{\perp}^{\mu} V_{\nu} \right) = 0, \label{eqVrot11} \\
&& k_{\parallel} \partial_z V_{\parallel} - f \partial_z (k_{\perp}^{\nu} V_\nu) = 0. \label{eqVrot12}
\end{eqnarray}

Here, the index \( \perp \) (parallel) denotes projection onto directions orthogonal (parallel) to the velocity \( u^{\mu} \). Projecting Eq.~(\ref{eqVrot11}) along the direction of \( u^\mu \), we find:
\begin{equation}\label{eq1rot2p}
zf \, \partial_z \left( z^{-1} \partial_z V_{\parallel} \right) + k_{\parallel} (k^{\nu}_{\perp} V_{\nu}) - k_{\perp}^2 V_{\parallel} = 0\,,
\end{equation}
while projection onto the transverse direction yields:
\begin{equation}\label{eq1rot2t}
zf \, \partial_z \left( z^{-1} f \, \partial_z V_{\perp}^{\alpha} \right) + (k_{\parallel}^2 - f k_{\perp}^2) V_{\perp}^{\alpha} + (f k^{\nu}_{\perp} V_{\nu} - k_{\parallel} V_{\parallel}) k_{\perp}^{\alpha} = 0\,.
\end{equation}

Thus, the system is composed of Eqs.~(\ref{eqVrot12}), (\ref{eq1rot2p}) and (\ref{eq1rot2t}). To eliminate unphysical redundancies, we introduce the gauge-invariant electric field~\cite{Morgan:2013dv}:
\begin{equation}\label{EqEV}
E_\alpha = F_{\alpha \nu} u^\nu = k_\alpha V_\parallel - k_\parallel V_\alpha.
\end{equation}
We then decompose the electric field into longitudinal and transverse parts with respect to the wave vector:
\begin{equation}\label{EqTL}
E^{\alpha}_{L} = \left( \frac{k^{\nu}_{\perp} E_{\nu}}{k_{\perp}^2} \right) k_{\perp}^\alpha, \quad
E^{\alpha}_{T} = \left( h^{\alpha \nu} - \frac{k^{\alpha}_{\perp} k^{\nu}_{\perp}}{k_{\perp}^2} \right) E_\nu.
\end{equation}

Substituting into equations (\ref{eqVrot12}), (\ref{eq1rot2p}) and (\ref{eq1rot2t}), we obtain:
\begin{equation}\label{EqLRot2}
\partial_z^{2} E_L^{\mu} + \left( -\frac{1}{z} + \frac{f'}{f} \frac{k^2_{\parallel}}{k^2_{\parallel} - f k^2_{\perp}} - \phi' \right) \partial_z E_L^{\mu} + \frac{k^2_{\parallel} - f k^2_{\perp}}{f^2} E_L^{\mu} = 0\,,
\end{equation}
\begin{equation}\label{EqTRot2}
\partial_z^{2} E_T^{\mu} + \left( -\frac{1}{z} + \frac{f'}{f} - \phi' \right) \partial_z E_T^{\mu} + \frac{k^2_{\parallel} - f k^2_{\perp}}{f^2} E_T^{\mu} = 0\,.
\end{equation}
Therefore, we decompose the equations of motion for heavy vector fields into two independent equations: one along the longitudinal direction and another along the transverse direction relative to the wave vector.

To study the effects of a rotating quark-gluon plasma  on quarkonium, we must  explicitly specify the wave vector on the boundary. For simplicity, we consider the following ansatz:
\begin{equation}
k^\mu = (-\omega, 0, 0, q)\,,
\end{equation}
where \( q \) is the momentum along the \( x^3 \) direction and \( \omega \) is the frequency. Consequently, the corresponding equation in terms of the longitudinal and transverse  electric field become:
\begin{equation}\label{EqLRot2_final}
\partial_z^{2} E_L^{\mu} + \left( -\frac{1}{z} + \frac{f'}{f} \frac{\gamma^2 \omega^2}{\gamma^2 \omega^2 - f(q^2 + \gamma^2 \omega^2 \bar{\Omega}^2)} - \phi' \right) \partial_z E_L^{\mu} + \frac{\gamma^2 \omega^2 - f(q^2 + \gamma^2 \omega^2 \bar{\Omega}^2)}{f^2} E_L^{\mu} = 0\,,
\end{equation}
\begin{equation}\label{EqTRot2_final}
\partial_z^{2} E_T^{\mu} + \left( -\frac{1}{z} + \frac{f'}{f} - \phi' \right) \partial_z E_T^{\mu} + \frac{\gamma^2 \omega^2 - f(q^2 + \gamma^2 \omega^2 \bar{\Omega}^2)}{f^2} E_T^{\mu} = 0\,.
\end{equation}
Note that these equations reduce to the zero-rotation case Refs.~\cite{Braga:2018hjt,Mamani:2013ssa} in the limit $\Omega \rightarrow 0$. 
Moreover,  in the presence of rotation, the temporal component of the electric field becomes nonzero.

\subsubsection{Boundary Conditions}

Let us now investigate the asymptotic behavior of the solutions of the gauge-invariant field equations (\ref{EqLRot2_final}) and (\ref{EqTRot2_final}). Near the horizon, the solution  is given by the infalling and ongoing  condition:
\begin{eqnarray}
 && E_{L/T}^{ \mu}(z\rightarrow z_h,k)=\left(1- \frac{z}{z_h}\right)^{-i \omega / (4 \pi T)}\left[ 1+a^{(1)}_{L/T}\left(1- \frac{z}{z_h}\right)+ \dots a^{(n)}_{L/T}\left((1- \frac{z}{z_h}\right)^{n}\right], \\ \label{Infalling} \cr && E_{L/T}^{ \mu}(z\rightarrow z_h,k)=\left(1- \frac{z}{z_h}\right)^{+i \omega / (4 \pi T)}\left[ 1+a^{(1)}_{L/T}\left(1- \frac{z}{z_h}\right)+ \dots a^{(n)}_{L/T}\left((1- \frac{z}{z_h}\right)^{n}\right], \label{ongoing}
\end{eqnarray}
where $a^{(1)}_{L/T}$,  $a^{(n)}_{L/T}$  are coefficients that depend on the frequency, momentum, and angular velocity.

On the other hand, near the boundary, the equations (\ref{EqLRot2_final}) and (\ref{EqTRot2_final}) can be solved in terms of normalizable and non-normalizable solutions, which exhibit the following asymptotic form:
\begin{equation}
 E_{L/T}^{ \mu}(z\rightarrow 0,k)=E_{0L/T}^{ \mu}+E_{2L/T}^{ \mu}z^2+\frac{1}{2}E_{0L/T}^{ \mu}\omega^2z^2\log z+\mathcal{O}(z^3)\,,
\end{equation}
where the coefficients $E_{0L/T}^{ \mu}$ and  $E_{2L/T}^{ \mu}$ are  dependent of the temperature, momentum and  frequency. This result is similar to the found in the case without rotation.

\section{Retarded Green's Function}\label{Retarded Green's Function}

\subsection{Retarded Green's Function}

The gauge theory current correlators can be described holographically by vector fields propagating in the bulk geometry, whose dynamics are governed by the action integral of Eq. (\ref{vectorfieldaction}), with the background metric given in Eq. (\ref{metricrot2}). In momentum space, the on-shell action can be expressed as
\begin{equation}\label{actionons}
S_{\text{on-shell}}=\frac{1}{2g^2_{5}}\int\frac{d\omega dq}{(2\pi)^2}e^{-\phi}\left[g^{zz}g^{\mu \nu}V_{\mu}(z,k)V_{\nu}'(z,k)\right].
\end{equation}

Upon substituting the relations provided by Eqs. (\ref{EqEV}) and (\ref{EqTL}) into Eq. (\ref{actionons}), we find:
\begin{eqnarray}\label{onshell2}
S_{\text{on-shell}}=\frac{1}{2g^2_{5}}\int\frac{d\omega dq}{(2\pi)^2}\frac{e^{-\phi}f}{k_{\parallel}^2z}\left[E_{T}^{\mu}(z,-k)\partial_zE_{T\mu}(z,k)+\frac{k^2_{\parallel}}{k_{\parallel}^2-k^2_{\perp}f}E_{L}^{\mu}(z,-k)\partial_zE_{L\mu}(z,k)\right]^{z_h}_0.
\end{eqnarray}
Next, we decompose the longitudinal and transverse fields in terms of their boundary values $E_{(0)L/T}^\mu$ as follows:
\begin{equation}\label{Bpropa}
E_{L/T}^{(-)\mu}(z,k)=\mathcal{E}^{\mu}_{L/T}(z,k)E_{(0)L/T}^\mu(k),
\end{equation}
with the functions $\mathcal{E}^{\mu}_{L/T}(z,k)$ satisfying the normalization condition $ \mathcal{E}^{\mu}_{L/T}(0,k)=1$. The superscript $(-)$ indicates that these gauge-invariant fields satisfy the infalling boundary condition, as required by the holographic prescription \cite{Son_2002}. Using this decomposition, Eq.~(\ref{onshell2}) transforms into
\begin{eqnarray}\label{onshell3}
    S_{on-shell}=&&\frac{1}{2g^2_{5}}\int\frac{d\omega dq}{(2\pi)^2}\frac{e^{-\phi}f}{k_{\parallel}^2z}\Bigg[E_{(0)T\nu}(-k)\mathcal{E}_{T}^{\mu}(z,-k)E_{(0)T}^\nu(k)\partial_z\mathcal{E}_{T\mu}(z,k) \nonumber \\ \cr +&&\frac{k^2_{\parallel}}{k_{\parallel}^2-k^2_{\perp}f}E_{(0)L}^\nu(-k)\mathcal{E}_{L}^{\mu}(z,-k)E_{(0)L\nu}(k)\partial_z\mathcal{E}_{L\mu}(z,k)\Bigg]^{z_h}_0
\end{eqnarray}
Following the Son-Starinets prescription \cite{Son_2002}, we rewrite the boundary values $E^{\mu}_{(0)L/T}(k)$  in terms of the original vector fields $V^{\mu}_{(0)}(k)$. Consequently, the on-shell action (\ref{onshell3}) can be compactly expressed as:
\begin{equation}\label{actioncompact}
S_{\text{on-shell}}=\frac{1}{2g^2_{5}}\int\frac{d\omega dq}{(2\pi)^2}\left[V_{\mu}^{(0)}(-k)\mathcal{F}^{\mu\nu}V^{(0)}_{\nu}(k)\right].
\end{equation}
with the components of $\mathcal{F}^{\mu\nu}$ explicitly given by:
\begin{eqnarray}\label{Fmunu}
    &&\mathcal{F}^{tt}=-\frac{e^{-\phi}f}{2g_{5}^{2}z}\left(\frac{q^4\gamma^2}{\left( \omega^2-q^2\right)(q^2+\gamma^2\omega^2\bar{\Omega}^2)}\mathcal{E}_{L}(z)\mathcal{E}_{L}'(z)+\frac{q^2\gamma^2\bar{\Omega}^2}{q^2+\gamma^2\omega^2\bar{\Omega}^2}\mathcal{E}_{T}(z)\mathcal{E}_{T}'(z)\right) \\  \cr && \mathcal{F}^{\varphi\varphi}=-\frac{e^{-\phi}f}{2g_{5}^{2}z}\left(\frac{(\omega^2-q^2)\bar{\Omega}^2\gamma^2}{(q^2+\gamma^2\omega^2\bar{\Omega}^2)}\mathcal{E}_{L}(z)\mathcal{E}_{L}'(z)+\frac{q^2\gamma^2}{q^2+\gamma^2\omega^2\bar{\Omega}^2}\mathcal{E}_{T}(z)\mathcal{E}_{T}'(z)\right) \\  \cr && \mathcal{F}^{x_3 x_3}=-\frac{e^{-\phi}f}{2g_{5}^{2}z}\left(\frac{q^3\gamma^2\omega}{\left( \omega^2-q^2\right)(q^2+\gamma^2\omega^2\bar{\Omega}^2)}\mathcal{E}_{L}(z)\mathcal{E}_{L}'(z)+\frac{q \omega\gamma^2 \bar{\Omega}^2}{q^2+\gamma^2\omega^2\bar{\Omega}^2}\mathcal{E}_{T}(z)\mathcal{E}_{T}'(z)\right) \\  \cr && \mathcal{F}^{x_2 x_2}=-\frac{e^{-\phi}f}{2g_{5}^{2}z}\mathcal{E}_{T}(z)\mathcal{E}_{T}'(z) \\  \cr && \mathcal{F}^{t \varphi }=- \mathcal{F}^{ \varphi t}=\frac{e^{-\phi}f}{2g_{5}^{2}z}\left(\frac{q^2\gamma^2\bar{\Omega}}{q^2+\gamma^2\omega^2\bar{\Omega}^2}\mathcal{E}_{L}(z)\mathcal{E}_{L}'(z)+\frac{q^2 \gamma^2 \bar{\Omega}}{q^2+\gamma^2\omega^2\bar{\Omega}^2}\mathcal{E}_{T}(z)\mathcal{E}_{T}'(z)\right)
    \\  \cr && \mathcal{F}^{t x_3 }=-\mathcal{F}^{x_3 t }=\frac{e^{-\phi}f}{2g_{5}^{2}z}\left(\frac{q^3\gamma^2\omega}{(\omega^2-q^2)(q^2+\gamma^2\omega^2\bar{\Omega}^2)}\mathcal{E}_{L}(z)\mathcal{E}_{L}'(z)+\frac{q \omega \gamma^2 \bar{\Omega}^2}{q^2+\gamma^2\omega^2\bar{\Omega}^2}\mathcal{E}_{T}(z)\mathcal{E}_{T}'(z)\right)  \\  \cr && \mathcal{F}^{ x_3 \varphi  }= -\mathcal{F}^{  \varphi  x_3 }=\frac{e^{-\phi}f}{2g_{5}^{2}z}\left(-\frac{q \omega \gamma^2 \bar{\Omega}}{q^2+\gamma^2\omega^2\bar{\Omega}^2}\mathcal{E}_{L}(z)\mathcal{E}_{L}'(z)+\frac{q \omega \gamma^2 \bar{\Omega}}{q^2+\gamma^2\omega^2\bar{\Omega}^2}\mathcal{E}_{T}(z)\mathcal{E}_{T}'(z)\right)
\end{eqnarray}
where the others components vanishing and  $\mathcal{E}_{L/T}$ are respectively the solution of the Eqs. (\ref{EqLRot2_final}) and (\ref{EqTRot2_final}) satisfying the infalling  boundary condition
\begin{eqnarray}\label{boundary2}
&& \mathcal{E}_{L/T}(z \rightarrow z_h,\omega) \longrightarrow \left(1-\frac{z}{z_h}\right)^{-i\omega/4\pi T}\,\left[1+a_{L/T}^{(1)}\left(1-\frac{z}{z_h}\right)+a_{L/T}^{(2)}\left(1-\frac{z}{z_h}\right)^2+...\right]\,,
\end{eqnarray}
and the normalization at the boundary,
\begin{equation}
\label{boundary2}
\mathcal{E}_{L/T}(z\to 0,\omega)=1\,.
\end{equation}
Finally, the two-point correlation function is obtained via the holographic prescription \cite{Son_2002}:
\begin{equation}
\label{Greens}
\mathcal{D}^{\mu \nu}=2\lim_{z \rightarrow0}\mathcal{F}^{\mu \nu}.
\end{equation}
Thus, the spectral function is extracted as:
\begin{equation}
\label{spectral}
\rho_{\mu \nu}=-Im\,D_{\mu \nu} .
\end{equation}

Although one could separately analyze each component of the spectral function, we simplify the calculations by projecting it onto the longitudinal and transverse directions relative to the wave vector. Employing the projection operators (\ref{EqTL})  we obtain
\begin{eqnarray}
&&\rho_{L} = \frac{k^{\alpha}k_{\perp}^\beta\rho_{\alpha\beta}}{k_{\perp}^2}= -Im  \left( \lim_{z\rightarrow 0}\frac{e^{-\phi}\gamma^2\omega^2f}{2g_{5}^{2}(q^2-\omega^2)z}\mathcal{E}_{L}(z)\mathcal{E}_{L}'(z)\right) \,, \label{eq:first}  \\ \cr && 
\rho_{T} = \left( h^{\alpha \beta} - \frac{k^{\alpha}_{\perp} k^{\beta}_{\perp}}{k_{\perp}^2} \right) \rho_{\alpha\beta}=Im  \left( \lim_{z\rightarrow 0}\frac{e^{-\phi}f}{2g_{5}^{2}z}\mathcal{E}_{T}(z)\mathcal{E}_{T}'(z)\right). \label{eq:second} 
\end{eqnarray}

\subsection{Numerical Results}

In order to analyze the behavior of charmonium in a rotating medium, it is necessary to fix the radius 
$l$ of the hyper-cylinder and select an angular velocity consistent with the phenomenology of non-central relativistic heavy-ion collisions.  Hydrodynamic simulations of heavy-ion collisions performed in \cite{Jiang:2016woz} predict even larger angular velocities, typically $\Omega \sim 20 - 40$ MeV. In case of 
Meanwhile, typical values for the QGP size have been estimated in Ref.~\cite{Sievert:2019zjr}, where the authors employed relativistic hydrodynamics to make predictions for potential future runs of ArAr and OO collisions at the Large Hadron Collider.
Based on these information, we set the radius to $l = 3$ fm and the angular velocity to $\Omega = 20$ MeV. This corresponds to  $\bar{\Omega} \sim 0.304$. 

\begin{figure}[!htb]
	\centering
	\includegraphics[scale=.45]{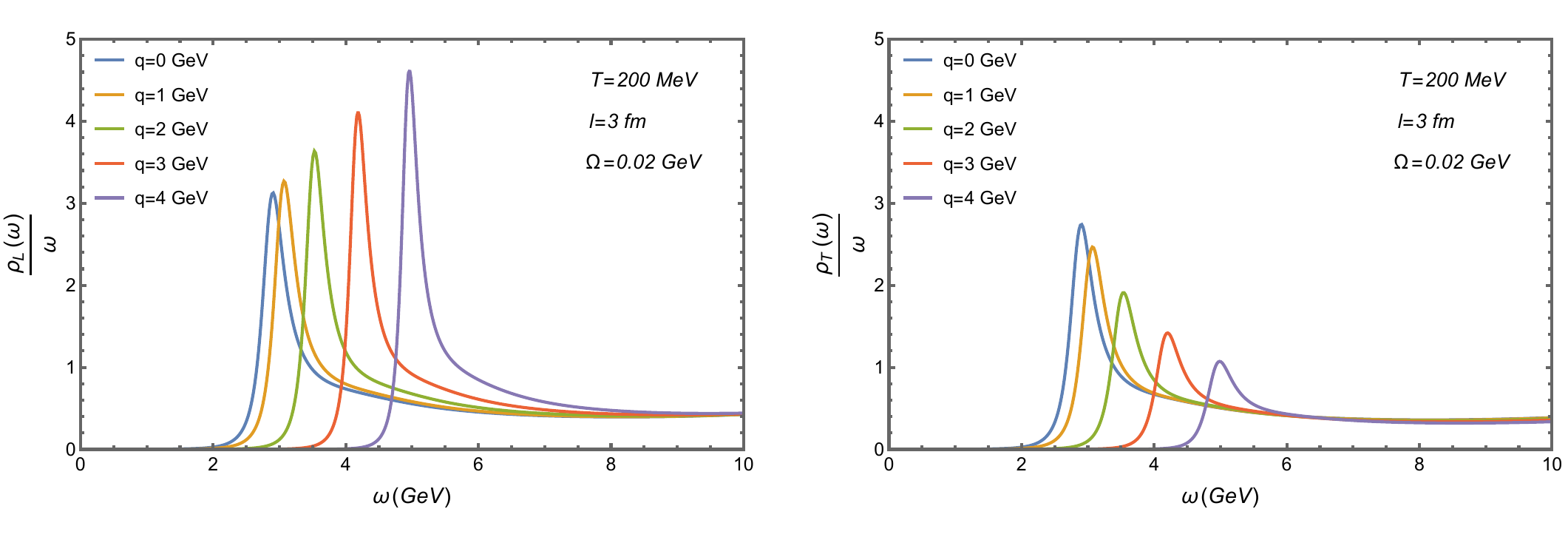}
	\caption{We display the spectral function for various values of linear momentum at $T = 200$ MeV, $\Omega = 20$ MeV, and $l = 3$ fm: the left panel corresponds to the longitudinal direction, while the right panel shows the transverse direction.}
    \label{Figspec1}
\end{figure}

The spectral functions for the transverse and longitudinal fields are obtained by numerically solving the respective equations (\ref{EqLRot2_final}) and (\ref{EqTRot2_final}), subject to the boundary conditions given by Eq. (\ref{boundary2}). Subsequently, the solutions are substituted into Eqs. (\ref{eq:first}) and (\ref{eq:second}) to obtain the final expressions. 
In Fig.~\ref{Figspec1}, we show the spectral functions for the transverse and longitudinal directions for various momentum values. One observes that the longitudinal peak becomes sharper and increases in height as momentum grows, indicating a reduction in the particle's width. Conversely, the transverse peak broadens significantly, suggesting that dissociation occurs more rapidly in the transverse sector.
It should be noted that, within the range of temperature and angular velocity analyzed, only the ground state remains. The excited states have already melted.

\begin{figure}[!htb]
	\centering
	\includegraphics[scale=.45]{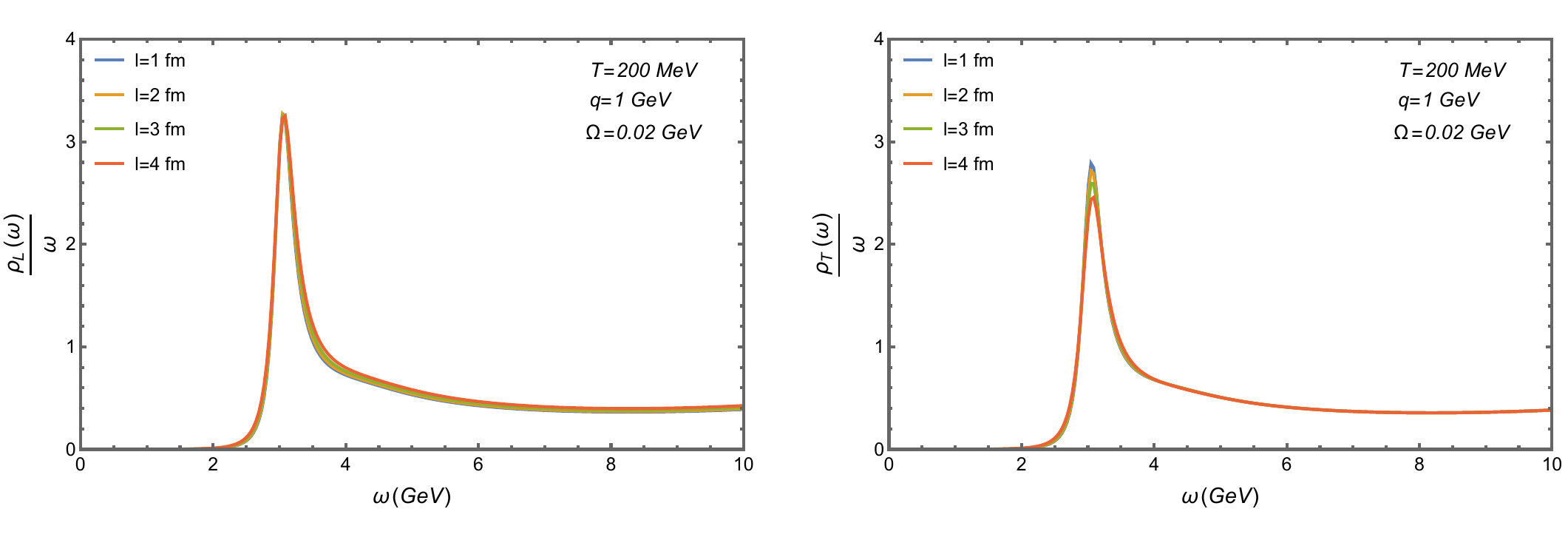}
	\caption{On the left, we show the spectral function in the longitudinal direction for various values of the radius $l$ at $q = 1$ GeV, $T = 200$ MeV, and $\Omega = 20$ MeV. On the right, the spectral function in the transverse direction is presented for the same set of parameters.}
    \label{Figspec2}
\end{figure}

One particularly interesting aspect to analyze is the variation of the rotational radius $l$. Since changes in the radius strongly affect the angular momentum, moment of inertia, and the overall properties of the rotating system, studying the spectral function as a function of $l$ may be relevant for future experimental investigations. Figure~\ref{Figspec2} shows the spectral functions at fixed angular velocity $\Omega = 20$ MeV, temperature, and linear momentum, but for different values of 
$l$. The results clearly demonstrate that variations in 
$l$ significantly influence the dissociation process. Notice that the effect is weaker in the longitudinal channel compared to the transverse one.

 \section{Quasinormal Modes}

Quasinormal modes represent the characteristic oscillations of black holes and black branes. Unlike many idealized macroscopic physical systems, perturbations in black hole spacetimes inherently involve dissipation due to the presence of an event horizon. Quasinormal modes typically have complex frequencies, where the imaginary part corresponds to the decay timescale of the perturbation, analogous to damping in classical physics. These eigenfunctions are generally non-normalizable and do not form a complete set.

Within the framework of gauge/gravity duality, quasinormal modes correspond to the poles of the retarded Green's functions for fluctuations around the black hole background, subject to infalling boundary conditions. In our case, these quasinormal modes represent quasi-particle states of vector mesons in the dual gauge theory. The real part of these modes corresponds to the mass of the vector mesons, while the imaginary part determines their decay rate.  

The peaks shown in Figures~\ref{Figspec1} and \ref{Figspec2} indicate that the corresponding retarded Green's functions possess poles, which are directly related to the frequencies of electromagnetic quasinormal modes of the black brane. In the dual gauge theory, these poles signify the presence of quasi-particle states of vector mesons. The frequencies of these QNMs, $\omega = \omega_{R} - i \omega_{I}$, have real and imaginary components: the real part reflects the mass, and the imaginary part corresponds to the decay rate of quasi-particle states.

Unlike the conformal AdS/CFT scenario (see, e.g., Refs.~\cite{Son:2002sd,Kovtun:2005ev}), where the temperature $T$ is the sole dimensionful parameter at finite temperature and quasinormal mode frequencies scale linearly with $T$, holographic QCD models demonstrate a more complex temperature dependence. In particular, the real part $\omega_R$ is associated with the quasi-state mass and, as $T \to 0$, converges to the vacuum mass of the corresponding hadronic state. The present holographic model incorporates three energy scales, allowing for simultaneous fits to zero-temperature masses and decay constants.

In the limit $T\rightarrow \infty$, the quasinormal  frequencies in holographic QCD models scale linearly with temperature, similar to the behavior observed in the conformal AdS case. However, this high-temperature regime corresponds to a region in which the mesonic states have already melted or dissociated in the thermal medium. Consequently, no peak appears in the spectral function.

In this section, we employ the shooting method, as described in Ref. \cite{Kaminski:2009ce}, to compute the quasinormal frequencies for electromagnetic perturbations in rotating plasma. This approach requires specifying boundary conditions at the horizon and subsequently adjusting a free parameter, the frequency, to satisfy the equations. Specifically, we solve equations (\ref{EqLRot2}) and (\ref{EqTRot2}):
\begin{equation}\label{EqLRot3}
    \partial_z^{2}E_L^{\mu}+\left( -\frac{1}{z}+\frac{f'}{f}\frac{\gamma^2\omega^2 }{\gamma^2\omega^2-f(q^2+\gamma^2\omega^2\bar{\Omega}^2)}-\phi' \right) \partial_z E_L^{\mu}+\frac{\left( \gamma^2\omega^2-f(q^2+\gamma^2\omega^2\bar{\Omega}^2)\right)}{f^2} E_L^{\mu}=0\,,
\end{equation}
\begin{equation}\label{EqTRot3}
   \partial_z^{2}E_T^{\mu}+\left( -\frac{1}{z}+\frac{f'}{f}-\phi' \right) \partial_z E_T^{\mu}+\frac{\left(  \gamma^2\omega^2-f(q^2+\gamma^2\omega^2\bar{\Omega}^2)\right)}{f^2} E_T^{\mu}=0\,,
\end{equation}
using the boundary conditions given by the infalling boundary conditions at the horizon position
\begin{equation}\label{infalling}
  \lim_{z\rightarrow z_h}  E_{L/T}^{ \mu}(z,k)=\left(1- \frac{z}{z_h}\right)^{-i\omega / (4 \pi T)}\left[ 1+a^{(1)}_{L/T}\left((1- \frac{z}{z_h}\right)+ \dots \right],
\end{equation}
and, the derivative of the infalling condition
\begin{eqnarray}
\lim_{z\rightarrow z_h}   \partial_z E_{L/T}^{ \mu}(z,k)&=&\left(1- \frac{z}{z_h}\right)^{-i \omega / (4 \pi T)}\left[ -\frac{a^{(1)}_{L/T}}{z_h}-\frac{a^{(2)}_{L/T}}{z_h}\left((1- \frac{z}{z_h}\right)+ \dots \right] \cr  &-& \frac{i \omega}{4 \pi T}\left(1-\frac{z}{z_h}\right)^{-1}\lim_{z\rightarrow z_h}E_{L/T}^{\mu}(z,p).
\end{eqnarray}
where
\begin{eqnarray}
   && a^{(1)}_{L}= 
\frac{z_h \left(4 i q^2 (\gamma  \omega  z_h+4 i)+\gamma ^2 \omega ^2 \left(-16 \bar{\Omega} ^2+i \gamma  \omega  \left(4 \bar{\Omega} ^2-3\right) z_h+2\right)-4 \gamma ^2 \omega ^2 z_h \phi '(z_h)\right)}{8 \gamma  \omega  (2i+\gamma  \omega  z_h)}\,,
\\ \cr &&a^{(1)}_{T}=
\frac{z_h \left(-4 \gamma  \omega  z_h \phi '(z_h)+i \left(4 q^2 z_h+\gamma  \omega  \left(\gamma  \omega  \left(4 \bar{\Omega} ^2-3\right) z_h-2 i\right)\right)\right)}{8 (2i+\gamma  \omega  z_h)}\,.
\end{eqnarray}
The coefficients involved in the infalling condition can be derived by inserting this condition into the equations of motion. Furthermore, it is essential to verify that the solution obtained from solving equations (\ref{EqLRot3}) and (\ref{EqTRot3}), using boundary conditions at the horizon, represents a quasinormal modes. This verification requires varying the frequency $\omega$  until the field satisfies the Dirichlet  conditions at the boundary.

\begin{figure}[!htb]
	\centering
	\includegraphics[scale=.45]{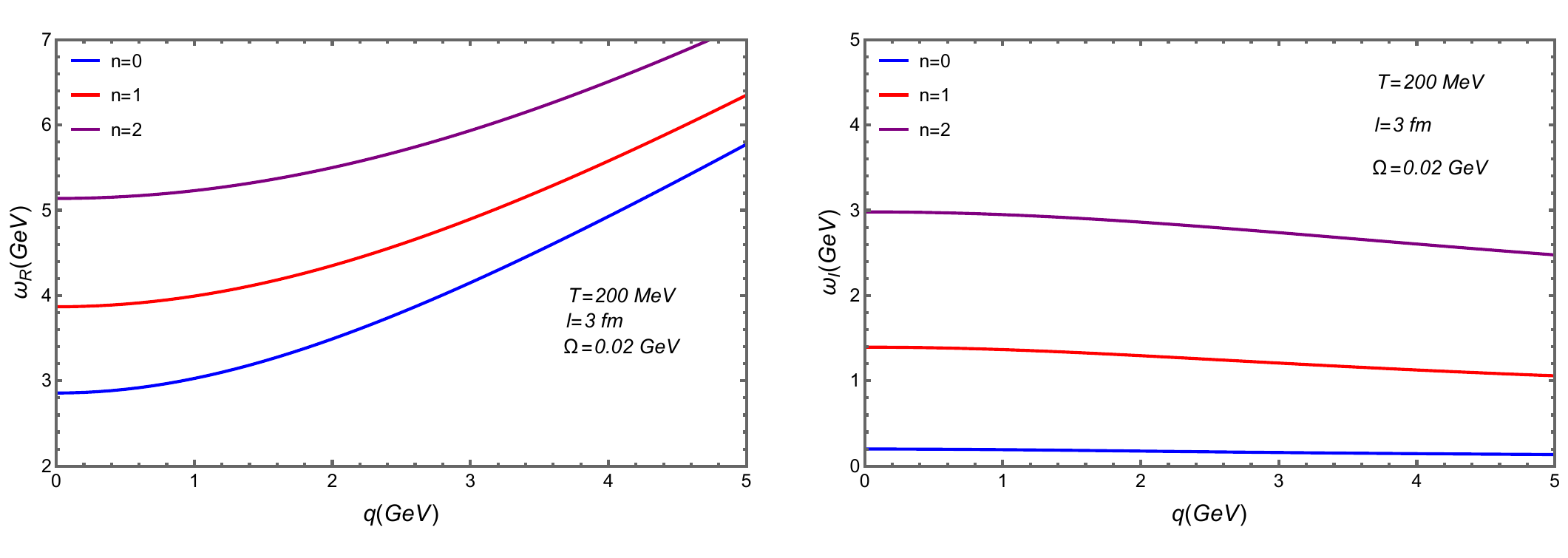}
	\caption{The frequencies of the first three quarkonium modes in the direction longitudinal to the wave vector, computed as a function of momentum using the shooting method at $T = 200$ MeV, $\Omega = 20$ MeV, and $l = 3$ fm, are presented.}
    \label{Fig1Lq}
\end{figure}

It is important to remark that in order to overcome the limitation of the shooting method hen large imaginary part $\omega_{I}$ are present,  it is necessary computing more coefficients of the near horizon expansion (\ref{infalling}) for higher temperatures in order to  find the quasinormal modes. This issue is discussed in \cite{Kaminski:2009ce}.

\subsection{Quasi-particles at the longitudinal sector}

We computed the quasinormal frequencies for charmonium states $n=0,1,2$ as a function of momentum using the shooting method to solve equation (\ref{EqLRot3}). The results are presented in Fig. \ref{Fig1Lq}, with the temperature fixed and the angular velocity specified. It is evident that as $q$ increases, the real part of the frequencies increases, while the imaginary part decreases. This behavior is similar to the results found in \cite{Braga:2018hjt}, where the rotation is not included. Notice from the analysis of the spectral function that the excited states have already dissociated. 

\begin{figure}[!htb]
	\centering
	\includegraphics[scale=.45]{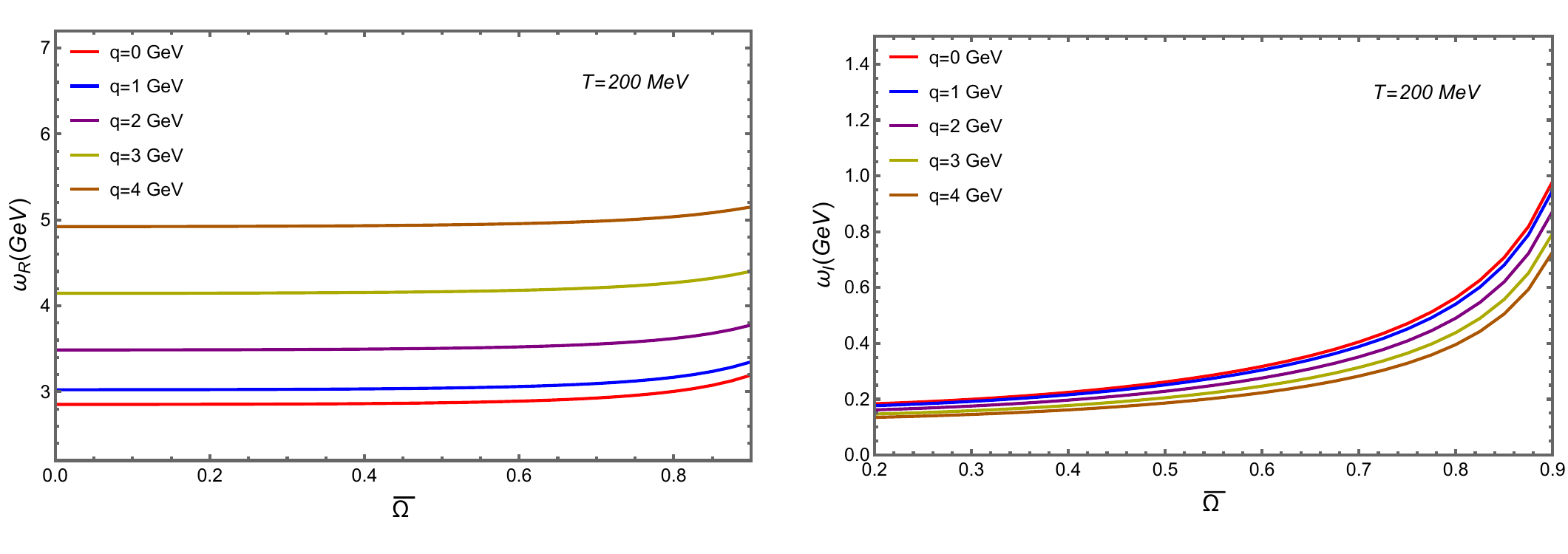}
	\caption{The quasinormal frequencies for the ground state are displayed as a function of $\bar{\Omega}$ for different values of momentum. The temperature is fixed at \( T = 200 \) MeV.}
    \label{Fig2Lqomega}
\end{figure}

In Fig. \ref{Fig2Lqomega}, we plot the ground state modes as a function of \( \bar{\Omega} \) for \( q = 0,1,2,3,4 \) GeV at a fixed temperature of \( T = 200 \) MeV. Notice that the frequencies increase rapidly as \( \bar{\Omega} \to 1 \), indicating that rotation induces the dissociation of the particles.  However, increasing momentum instead decreases the width, suggesting that higher momentum suppresses the dissociation of particles in the longitudinal direction. This is also shown in the analyze of the spectral function as can be seen in Fig. \ref{Figspec1}.

\begin{figure}[!htb]
	\centering
	\includegraphics[scale=.70]{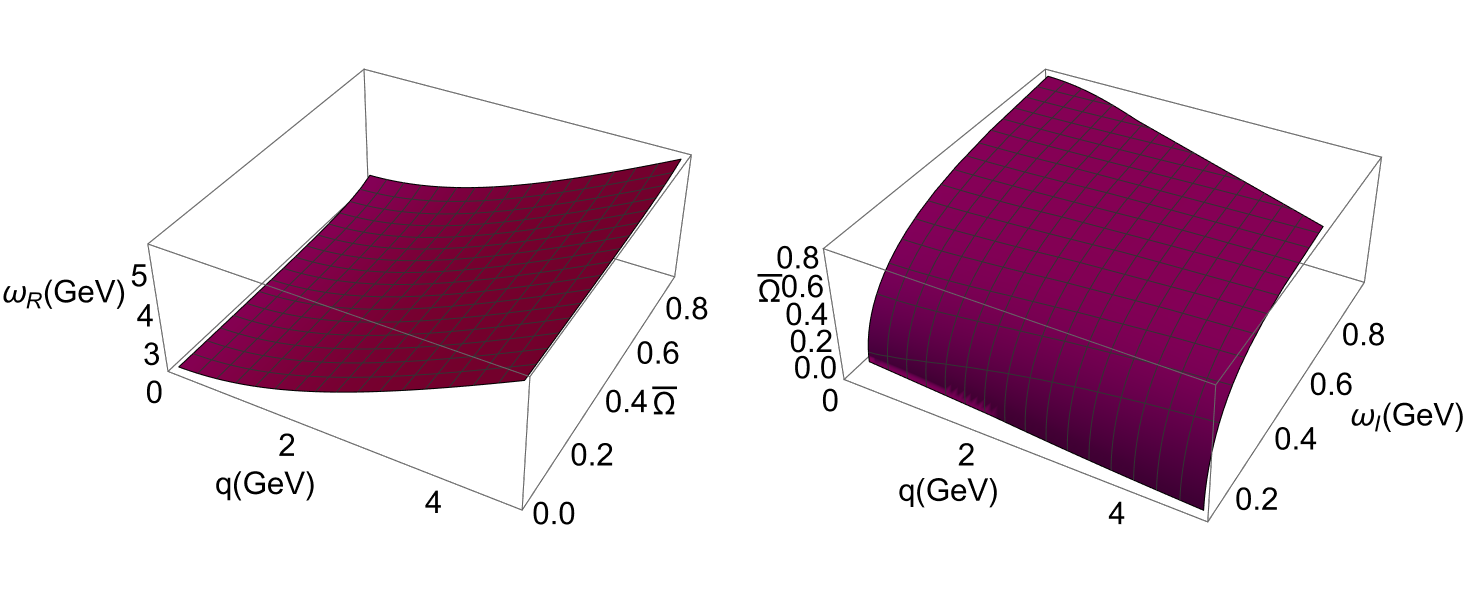}
	\caption{3D plots of the quasinormal frequencies for the ground state at $200$ MeV. The right panel shows the real part of the frequency as a function of momentum and angular velocity, while the left panel shows the imaginary part.}
    \label{Fig3Lqomega}
\end{figure}

To illustrate these observations, we present a 3D plot of the momentum, angular velocity, and quasinormal frequencies of the ground state in Fig. \ref{Fig3Lqomega}. The figure in the right  shows the values of the real part of the frequency, while the figure in the left presents the results for the imaginary part.

In addition to the modes discussed above, the longitudinal channel features an additional mode associated with hydrodynamic behavior. Since the pole-skipping phenomenon emerges in hydrodynamic modes, we will analyze it in detail in the following section.

\subsubsection{Hydrodynamic Limit}

The gauge/gravity duality provides a powerful framework to study additional physical properties of deconfined plasmas, such as the computation of transport coefficients in the hydrodynamic regime. This regime is characterized by low frequencies and small wave numbers compared to the temperature. By analyzing the behavior of field perturbations in this limit, one can extract the diffusion constant using established techniques for evaluating hydrodynamic transport coefficients~\cite{Policastro:2002se,Policastro:2002tn,Ballon-Bayona:2024twa}.

To compute the diffusion constant, it is convenient to normalize the relevant quantities by the temperature, thereby rendering both the rescaled variables and the holographic coordinate dimensionless:
\begin{eqnarray}
     \mathfrak{w}=\frac{\omega}{\gamma\pi T}, \qquad \mathfrak{q}=\frac{q}{\gamma\pi T}, \qquad u=(\gamma\pi T)z\,.
\end{eqnarray}
In terms of these dimensionless variables, Eq.~(\ref{EqLRot3}) becomes:
\begin{equation}\label{eqdimensionless}
\partial_{u}^2 E_{L}^{\mu}+\left( -\frac{1}{u}+\frac{f'}{f}\frac{ \mathfrak{w}^2}{\mathfrak{w}^2-f(\mathfrak{q}^2+\gamma^2\mathfrak{w}^2\bar{\Omega}^2)}-\phi' \right) \partial_{u}E_{L}^{\mu}+\frac{\left(\mathfrak{w}^2-f(\mathfrak{q}^2+\gamma^2\mathfrak{w}^2\bar{\Omega}^2)\right)}{f^2} E_{L}^{\mu}=0\,.
\end{equation}

To implement ingoing boundary conditions at the horizon, we perform the following field redefinition:
\begin{equation}\label{EqRe}
    E_{L}^{\mu}(u)=f^{-\frac{i\mathfrak{w}\gamma}{4}}Y_{L}^{\mu}(u)\,,
\end{equation}
where \( Y_L^{\mu}(u) \) is a regular function at the horizon. Substituting this expression into Eq.~(\ref{eqdimensionless}), we obtain the following second-order differential equation for \( Y_L^{\mu} \):
\begin{eqnarray}\label{eqdimensionless2}
&& \partial_{u}^{2}Y_{L}^{\mu} - \partial_{u}Y_{L}^{\mu}
    \left[ \frac{1}{u} + \frac{i\gamma \mathfrak{w} f'}{2f} + \phi' 
    + \frac{\gamma^2 \mathfrak{w}^2 f'}{f \left( -\gamma^2 \mathfrak{w}^2 + f\left(\mathfrak{q}^2 + \gamma^2 \mathfrak{w}^2 \bar{\Omega}^2 \right)\right)} 
    \right] \nonumber \\ 
&& \quad + Y_{L}^{\mu}
    \left[ -\frac{i\gamma \mathfrak{w} f''}{4f} 
    + \frac{\gamma^2 \mathfrak{w}^2 - f\left(\mathfrak{q}^2 - \gamma^2 \mathfrak{w}^2 \bar{\Omega}^2 \right)}{f^2} 
    + \frac{i\gamma \mathfrak{w} f'}{4f} 
    \left( \frac{1}{u} + \phi' + \frac{\gamma^2 \mathfrak{w}^2 f'}{f\left( -\gamma^2 \mathfrak{w}^2 + f\left(\mathfrak{q}^2 + \gamma^2 \mathfrak{w}^2 \bar{\Omega}^2 \right)\right)} \right) \right. \nonumber \\ 
&& \quad \left. -\frac{\gamma \mathfrak{w} \left(\gamma \mathfrak{w} - 4i\right) f'^2}{16f^2} \right] = 0.
\end{eqnarray}

In the hydrodynamic limit, where both the frequency and the wave number are much smaller than the temperature (i.e., \( \mathfrak{w} \ll 1 \), \( \mathfrak{q} \ll 1 \)), it is customary to develop perturbative solutions. Instead of using the standard multi-parameter expansion, we adopt an alternative approach introduced in Ref.~\cite{Mamani:2022qnf}. We rescale the small parameters as follows:
\[
\mathfrak{w} \rightarrow \lambda \mathfrak{w}, \quad \bar{\Omega} \rightarrow \lambda \bar{\Omega}, \quad \mathfrak{q} \rightarrow \lambda \mathfrak{q}, \quad \lambda \ll 1,
\]
and expand the solution in powers of \( \lambda \):
\begin{equation}\label{ExpansionHydro}
    Y_{L}^{\mu}(u)=Y_{(0)}^{\mu}(u)+\lambda Y_{(1)}^{\mu}(u)+\lambda^2 Y_{(2)}^{\mu}(u)+\dots\,.
\end{equation}
By inserting this expansion into Eq.~(\ref{eqdimensionless2}) and solving order by order under regularity at the horizon, we obtain:
\begin{eqnarray}
&& Y_{(0)}^{\mu}(u) = Y_0^{\mu}, \nonumber \\
&& Y_{(1)}^{\mu}(u) = iY_0^{\mu} \frac{\gamma \mathfrak{w}}{4\pi T} \ln f
- \frac{i Y_0^{\mu} \mathfrak{q}^2}{\gamma \mathfrak{w}} \frac{e^{-\phi(u_h)} f'(u_h)}{4\pi T u_h} \left( \int_0^{u_h} dx\, x e^{\phi} - \int_0^u dx\, x e^{\phi} \right) \nonumber \\
&& \quad - iY_0^{\mu} \gamma \mathfrak{w} \frac{e^{-\phi(u_h)} f'(u_h)}{4\pi T u_h} \int_0^u dx\, \frac{x e^{\phi}}{f}\,.
\end{eqnarray}
Substituting the solution into Eq.~(\ref{EqRe}) and imposing the Dirichlet boundary condition \( E_L^{\mu}(0) = 0 \), we find the dispersion relation:
\begin{equation}
\omega = \frac{i}{\gamma} \left[ \frac{e^{-\phi(u_h)} f'(u_h)}{4\pi T u_h} \int_0^{u_h} dx\, x e^{\phi} \right] q^2\,.
\end{equation}
Comparing this to Fick’s law, \( \omega = -i D q^2 \), we get the diffusion coefficient
\begin{equation}
D = - \frac{1}{\gamma} \frac{e^{-\phi(u_h)} f'(u_h)}{4\pi T u_h} \int_0^{u_h} dx\, x e^{\phi}\,.
\end{equation}
Finally, this can be written in terms of the original coordinate as:
\begin{equation}\label{Diffzh}
D = \frac{e^{-\phi(z_h)}}{z_h} \int_0^{z_h} dz\, z e^{\phi(z)}\,.
\end{equation}
Although Eq.~(\ref{Diffzh}) does not explicitly depend on the angular velocity, such dependence enters indirectly through the horizon position \( z_h \), which is related to the temperature via Eq.~(\ref{temp}) and is itself a function of \( \bar{\Omega} \). To illustrate this, consider the AdS\(_5\) case with \( \phi = 0 \). Evaluating Eq.~(\ref{Diffzh}) gives:
\begin{equation}\label{DAdS}
D_{\mathcal{N}=4\, \text{SYM}} = \frac{1}{2 \gamma \pi T}\,,
\end{equation}
showing explicitly that the effect of rotation is encoded in the Lorentz factor \( \gamma \). The remaining part of the expression corresponds to the result in the non-rotating case.

\begin{figure}[!htb]
\centering
\includegraphics[scale=.5]{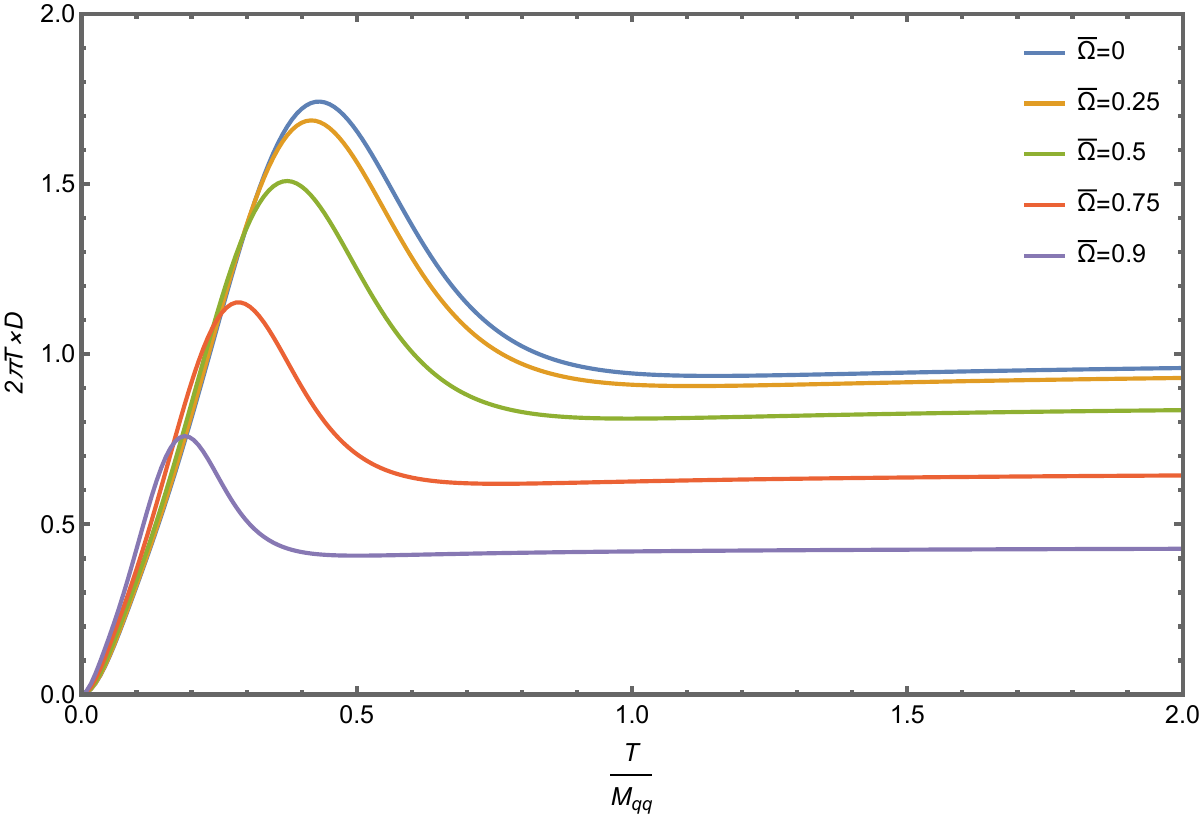}
\caption{Numerical results for the charge diffusion coefficient of charmonium at various angular velocity values, scaled by \( 2\pi T \), as a function of temperature rescaled by the mass of the first hadronic state.}
\label{Diffusion}
\end{figure}

For the dilaton profile considered here, the integral in Eq.~(\ref{Diffzh}) cannot be solved analytically and must be computed numerically. In Fig.~\ref{Diffusion}, we present the results for the product \( 2\pi T \times D \) as a function of temperature, scaled by the mass of the first hadronic state. At low temperatures, a non-trivial dependence on \( T \) is observed, while at high temperatures the diffusion coefficient approaches the AdS/CFT value: \( D_{\text{Rot}} = 1/(2\gamma \pi T) \). Thus, in the conformal limit, the sole effect of rotation is the appearance of the  factor \( \gamma \).

\subsection{Quasi-particles at the transverse sector}

\begin{figure}[!htb]
	\centering
	\includegraphics[scale=.42]{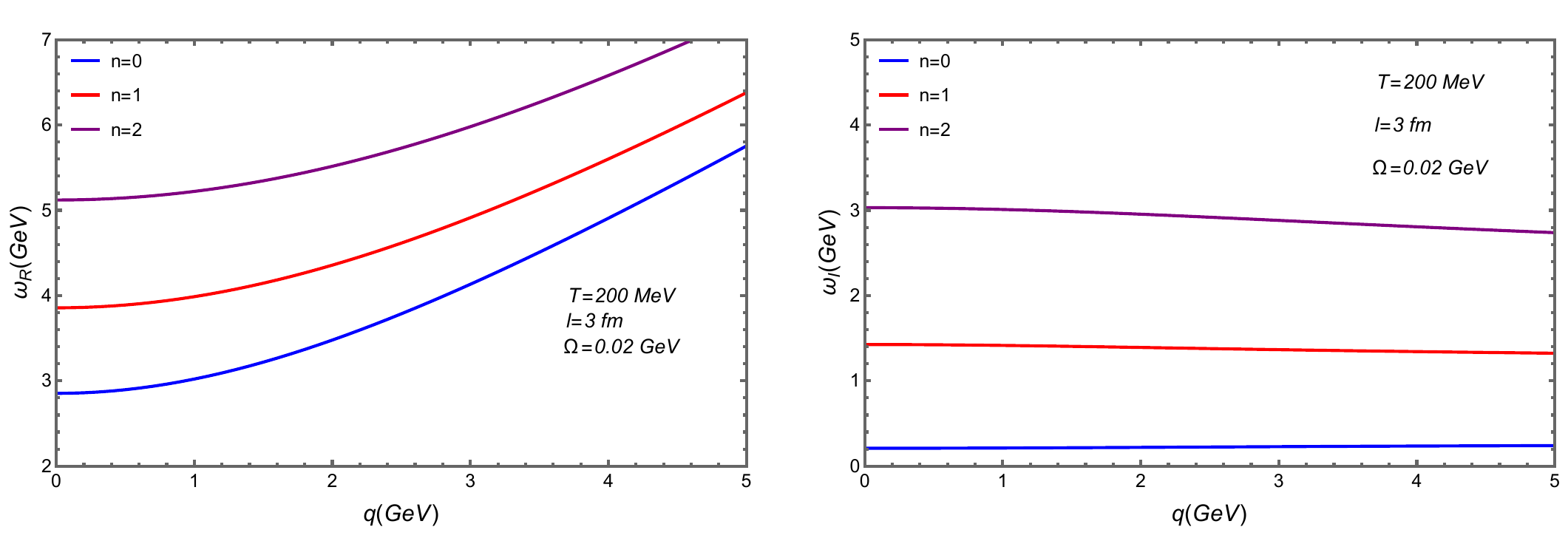}
	\caption{The frequencies of the first three quarkonium modes in the direction trasnverse to the wave vector, computed as a function of momentum using the shooting method at $T = 200$ MeV, $\Omega = 20$ MeV, and $l = 3$ fm, are presented.}
    \label{Fig1Tq}
\end{figure}

In the transverse sector, the quasinormal modes are computed by solving Eq.~(\ref{EqTRot3}) using the shooting method. The results for the quasinormal frequencies of the first three modes, \( n = 0, 1, 2 \), are shown in Fig.~\ref{Fig1Tq} as a function of momentum, with fixed temperature \( T = 200 \) MeV and angular velocity \( \bar{\Omega} = 0.304 \). The real part of the quasinormal frequencies exhibits behavior similar to that of the longitudinal sector. However, the imaginary part behaves differently: in the transverse sector, the width increases compared to the longitudinal case. These results suggest that meson dissociation is more pronounced in directions transverse to the wave vector. As in the longitudinal case, the excited states have already melted, and their widths are extremely large.

\begin{figure}[!htb]
	\includegraphics[scale=.45]{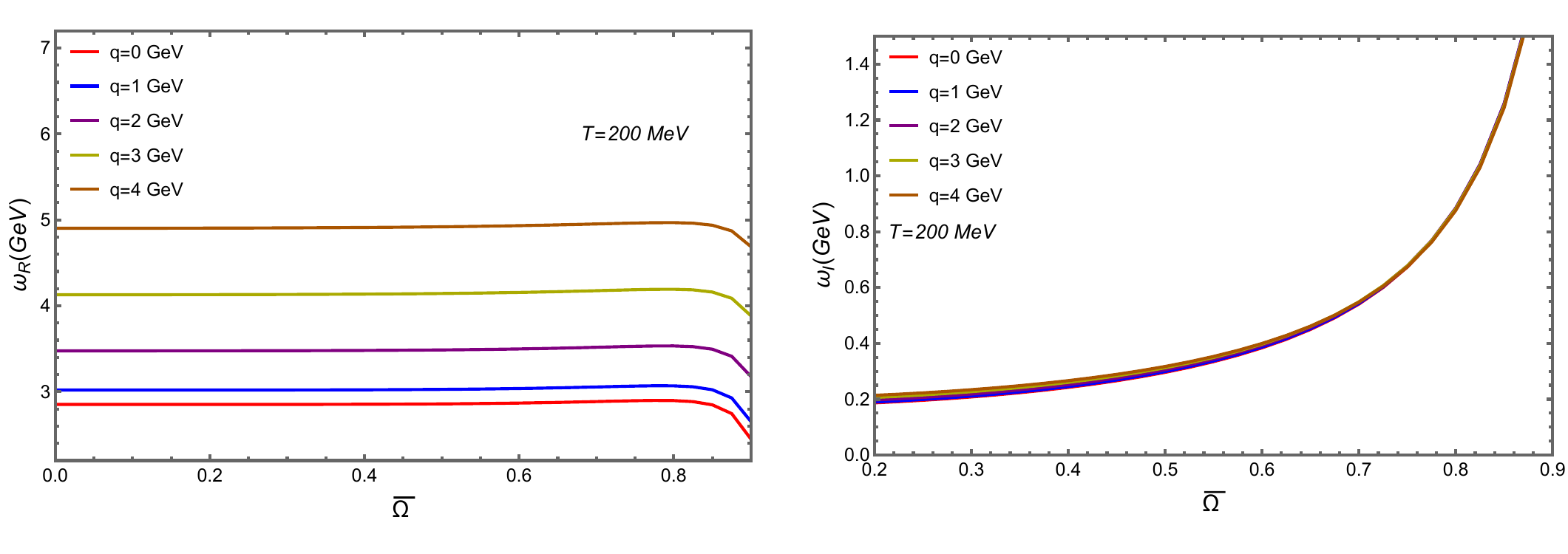}
	\caption{Quasinormal frequencies for the ground state as a function of $\bar{\Omega}$, for different values of momentum. The temperature is fixed at \( T = 200 \) MeV.}
    \label{Fig2Tqomega}
\end{figure}

Figure~\ref{Fig2Tqomega} shows the behavior of the quasinormal modes for the ground state at different momenta as the angular velocity varies, with temperature kept constant. In contrast to the longitudinal sector, the real part of the transverse quasinormal frequencies decreases as \( \bar{\Omega} \to 1 \). The imaginary part also behaves differently: the width decreases with increasing momentum. 

\begin{figure}[!htb]
	\centering
	\includegraphics[scale=.65]{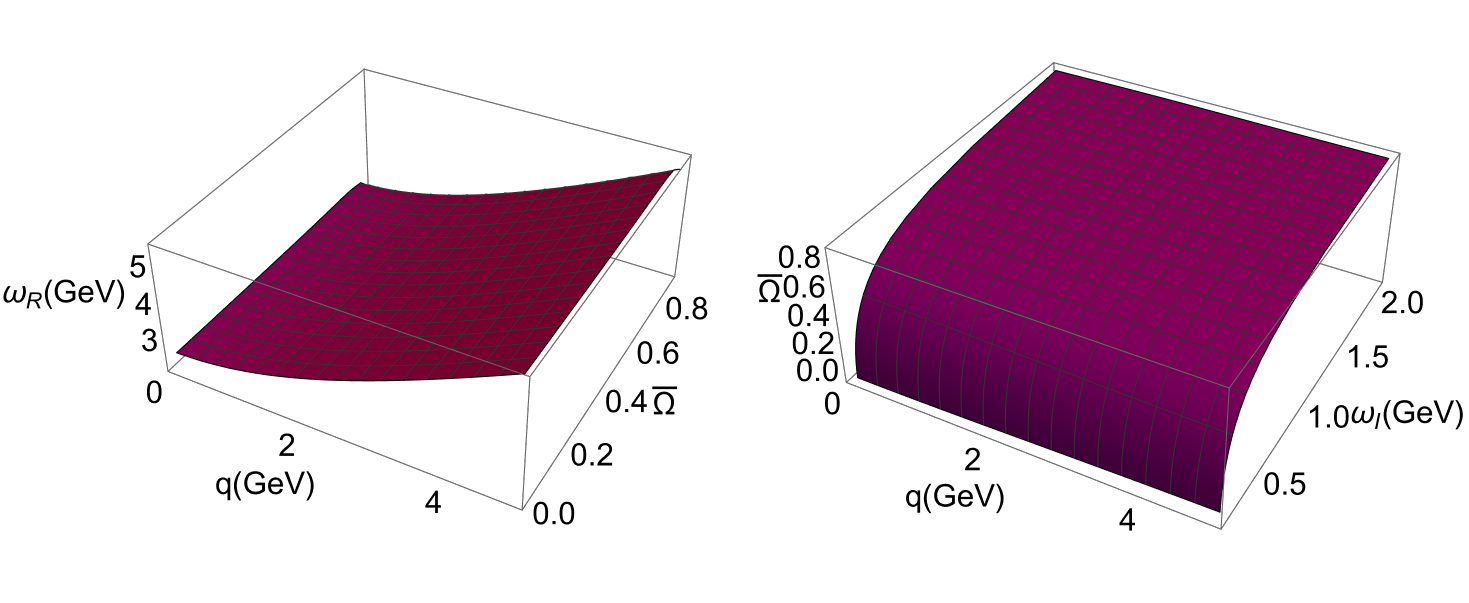}
	\caption{3D plots of the quasinormal frequencies for the ground state at $200$ MeV. The right panel shows the real part of the frequency as a function of momentum and angular velocity, while the left panel shows the imaginary part.}
    \label{Fig3Tqomega}
\end{figure}

As in the longitudinal case, we present a 3D plot in Fig.~\ref{Fig3Tqomega} to further illustrate the dependence of the ground state quasinormal modes on both momentum and angular velocity in the rotating background. The right panel displays the real part of the frequency, while the left panel shows the imaginary part.

To compare the spectral functions of the longitudinal and transverse sectors with quasinormal modes, we use the Breit-Wigner distribution to extract the thermal mass and width on the spectral function presented in \ref{Retarded Green's Function}. The Breit-Wigner expression is given by \cite{Miranda:2009uw,Colangelo:2009ra}:
\begin{equation}\label{BWdistri}
\mathcal{R}(\omega)=\frac{a\omega^{b}}{(\omega-\omega_R)^2+\omega_I^2},
\end{equation}
where $\omega_I$ is the  width and $\omega_R$ is the real part of the frequency. The quantities $a$ and $b$ are adjustable constants that depend on the temperature, $\bar{\Omega}$, momentum, and the position of the peak along the frequency axis. Table \ref{TabBW} presents the results for Fig. \ref{Figspec1}, demonstrating consistency with the spectral function analysis. 
\begin{table}[htbp]
  \centering
  \begin{tabular}{||c||c|c||c|c||c||}
    \hline
    \multicolumn{1}{||c||}{Longitudinal} 
    & \multicolumn{2}{|c||}{Breit-Wigner} 
    & \multicolumn{2}{|c||}{QNMS} 
    \\
    \cline{2-5}
    \multicolumn{1}{||c||}{$q(GeV)$} 
    & $\omega_R$ (GeV) & $\omega_I$ (GeV) 
    & $\omega_R$ (GeV) & $\omega_I$ (GeV)  \\
    \hline
    $0$ & $2.856$ & $2.021 \times 10^{-3}$ & $2.857$ & $2.001 \times 10^{-3}$  \\
    $1$ & $3.027$ & $1.943 \times 10^{-3}$ & $3.028$ & $1.929 \times 10^{-3}$  \\
    $2$ & $3.491$ & $1.769 \times 10^{-3}$ & $3.491$ & $1.761 \times 10^{-3}$  \\
    $3$ & $4.151$ & $1.598 \times 10^{-3}$  & $4.151$ & $1.594\times 10^{-3}$  \\  $4$ & $4.928$ & $1.466 \times 10^{-3}$  & $4.927$ & $1.46 \times 10^{-3}$  \\
    \hline
    \multicolumn{1}{||c||}{Transverse} 
    & \multicolumn{2}{|c||}{Breit-Wigner} 
    & \multicolumn{2}{|c||}{QNMS} 
    \\
    \cline{2-5}
     \multicolumn{1}{||c||}{$q(GeV)$} 
    & $\omega_R$ (GeV) & $\omega_I$ (GeV) 
    & $\omega_R$ (GeV) & $\omega_I$ (GeV)  \\
    \hline
    $0$ & $2.851$ & $2.109 \times 10^{-3}$ & $2.853$ & $2.092 \times 10^{-3}$  \\
    $1$ & $3.019$ & $2.138 \times 10^{-3}$ & $3.020$ & $2.122 \times 10^{-3}$  \\
    $2$ & $3.477$ & $2.206 \times 10^{-3}$ & $3.485$ & $2.152 \times 10^{-3}$  \\
    $3$ & $4.133$ & $2.289\times 10^{-3}$  & $4.132$ & $2.280\times 10^{-3}$  \\  $4$ & $4.908$ & $2.366 \times 10^{-3}$  & $4.907$ & $2.352 \times 10^{-3}$  \\
    \hline
  \end{tabular}
  \caption{The ground state frequencies, obtained both from the Breit-Wigner distribution (see Eq.~\ref{BWdistri}) and from the QNM calculation, are presented as a function of momentum for $T = 200$ MeV, $\Omega = 0.020$ GeV, and $l = 3$ fm.}
  \label{TabBW}
\end{table}

\section{Pole-Skipping in the Charmonium Model}

Recently, the phenomenon of pole-skipping has been identified in holographic Green’s functions \cite{Grozdanov:2017ajz, Blake:2017ris, Blake:2018leo}. Originally observed in studies of quantum many-body chaos \cite{Grozdanov:2017ajz, Blake:2017ris, Blake:2018leo}, pole-skipping refers to special points in momentum space where the retarded two-point Green’s function becomes ill-defined. However, pole-skipping is not exclusive to quantum chaos; an infinite set of pole-skipping points exists, some of which are unrelated to chaos. To systematically identify these points, a holographic near-horizon method was developed in Ref. \cite{Blake:2019otz}. This method has been applied in various holographic models, as discussed in Refs. \cite{Natsuume:2019sfp, Natsuume:2019xcy, Ahn:2019rnq, Ahn:2020baf, Wang:2022mcq, Pan:2024azf}.  Additional studies addressing the effects of rotation are presented in Refs.~\cite{Blake:2021hjj,Jeong:2023rck}.

\begin{figure}[H]
	\centering
	\includegraphics[scale=.45]{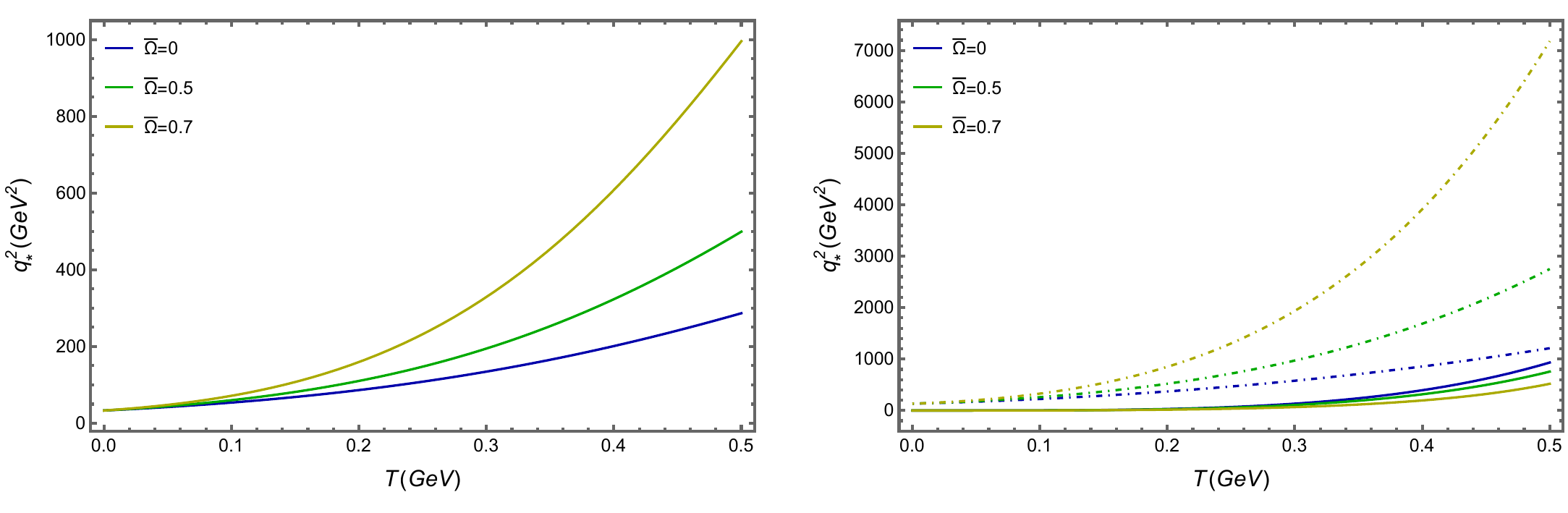}
	\caption{Pole-skipping values of the squared momentum as a function of temperature for the longitudinal case, computed for different values of $\bar{\Omega}$. The left panel shows the results for the first-order pole-skipping, while the right panel presents the results for the second order. In the right panel, the dot-dashed line represents the squared momentum $q_{*+}^2$, whereas the solid line corresponds to $q_{*-}^2$.}
    \label{PoleskippingsL}
\end{figure}

This phenomenon also appears in the holographic Green’s function of the charmonium model, making the identification of pole-skipping points a matter of particular significance. The procedure for determining the pole-skipping locations in both the longitudinal and transverse sectors follows the methodology presented in Appendix~\ref{A1}, which is based on the approach developed in Ref.~\cite{Blake:2019otz}. 

In the longitudinal sector, the first three pole-skipping points are given by:
\begin{align}\label{poleLong}
   &\text{1st order:} 
   \left\{ 
   \begin{aligned} 
       \omega_{*} &= 0;  \\  
       q_{*}^{2} &= 0;  
   \end{aligned} 
   \right.
   \notag \\[10pt]
   &\text{2nd order:} 
   \left\{ 
   \begin{aligned} 
       \omega_{*} &= -\frac{2i}{z_h\gamma}=-2i\pi T;  \\  
       q_{*}^{2} &= \frac{2 + 4\bar{\Omega}^2 +2z_h\phi'(z_h)}{z_h^2};  
   \end{aligned} 
   \right.
   \notag \\[10pt]
   &\text{3rd order:} 
   \left\{ 
   \begin{aligned} 
       \omega_{*} &= -\frac{4i}{z_h\gamma}=-4i\pi T;  \\  
       q_{*\pm}^2 &= \frac{2 \left( -2 + 8 \bar{\Omega}^2 + z_h \phi'(z_h) \pm \sqrt{12 + z_h \left( \phi'(z_h) \left( 8 + z_h \phi'(z_h) \right) + 4 z_h \phi''(z_h) \right)} \right)}{z_h^2};  
   \end{aligned} 
   \right.
\end{align}

\begin{figure}[H]
	\centering
	\includegraphics[scale=.4]{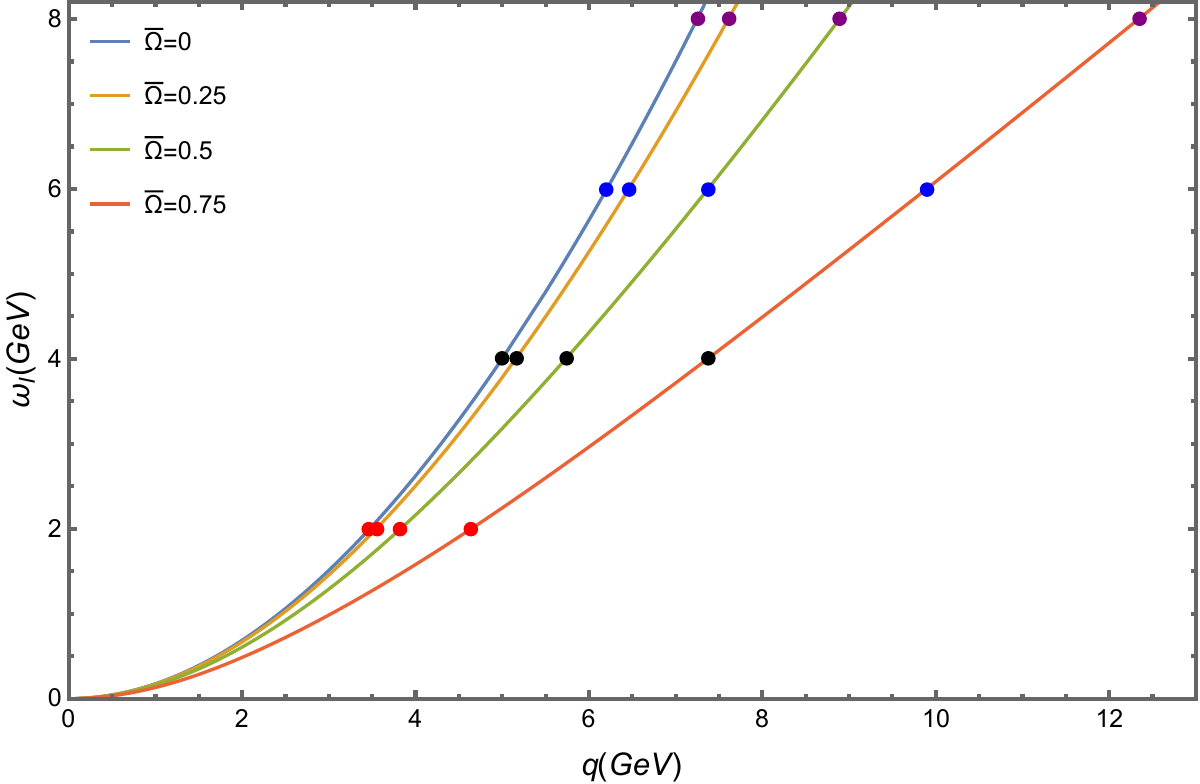}
	\caption{Hydrodynamical modes found in the longitudinal sector for various values of $\bar{\Omega}$ at $T=1/\pi$ GeV, where the red points correspond to second-order pole-skipping in the mode, the black points represent third-order pole-skipping, the blue points indicate fourth-order pole-skipping, and the purple points correspond to fifth-order pole-skipping.}
    \label{Poleskippings}
\end{figure}

Figure~\ref{PoleskippingsL} displays the pole-skipping points of the squared momentum in the longitudinal sector as a function of temperature, computed for different values of the angular velocity. Note that the momentum values associated with these pole-skipping points are always positive. These points are associated with the hydrodynamic modes. To illustrate this connection, in Fig.~\ref{Poleskippings} we present the hydrodynamic modes in the longitudinal channel for several values of the angular velocity. The corresponding pole-skipping points located along these modes are indicated by colored dots. Specifically, the second-order pole-skipping point is represented in red, the third-order in black, the fourth-order in blue, and the fifth-order in purple. These values were obtained both analytically, from Eq.~(\ref{poleLong}), and numerically. From Fig.~\ref{Poleskippings}, it can also be observed that increasing the angular velocity results in a corresponding increase in the momentum values.

In the transverse sector, the pole-skipping  points found are 
\begin{align}\label{firstorder}
   &\text{1st order:} 
   \left\{ 
   \begin{aligned} 
       \omega_{*} &= -\frac{2i}{z_h\gamma}=-2i\pi T;  \\  
       q_{*}^2 &= \frac{-2 + 4\bar{\Omega}^2 - 2z_h\phi'(z_h)}{z_h^2};  
   \end{aligned} 
   \right.
   \notag \\[10pt]
   &\text{2nd order:} 
   \left\{ 
   \begin{aligned} 
       \omega_{*} &= -\frac{4i}{z_h\gamma}=-4i\pi T;  \\  
       q_{*\pm}^2 &= \frac{2 \left( -4 + 8 \bar{\Omega}^2 -z_h \phi'(z_h) \pm \sqrt{8 + z_h \left( \phi'(z_h) \left( -4 + z_h \phi'(z_h) \right) - 4 z_h \phi''(z_h) \right)} \right)}{z_h^2};  
   \end{aligned} 
   \right.
   \notag \\[10pt]
   &\text{3rd order:} 
   \left\{ 
   \begin{aligned} 
       \omega_{*} &= -\frac{6i}{z_h\gamma}=-6i\pi T;  \\  
       q_{*\pm}^2 &= - \frac{2 \left(-18 \Omega^2 + z_h \phi'(z_h) + 9 \right)}{z_h^2} \pm\frac{2i(\sqrt{3} \pm i)\sqrt[3]{\frac{2}{3}}\, A}{z_h^2 \sqrt[3]{B_1 + B_2}}
\mp \frac{i(\sqrt{3}+i)\left(\frac{2}{3}\right)^{2/3} \sqrt[3]{B_1+ B_2}}{z_h^2}; \\   q_{*}^{2}&=- \frac{2 \left(-18 \Omega^2 + z_h \phi'(z_h) + 9 \right)}{z_h^2} + \frac{4 \left(\frac{2}{3}\right)^{1/3} \, A}{z_h^2 (B_1+ B_2)^{1/3}}
+ \frac{4\left(\frac{2}{3}\right)^{2/3}(B_1+ B_2)^{1/3}}{z_h^2};  
   \end{aligned} 
   \right.
\end{align}
with
\begin{eqnarray}
  &&  A=-4 z_h^2 \phi''(z_h) + z_h^2 \phi'(z_h)^2 - 4 z_h \phi'(z_h) + 14 \,;\nonumber \\ \cr && B_1=
-27 z_h^3 \phi^{(3)}(z_h) + 18 z_h^3 \phi'(z_h) \phi''(z_h) - 81 z_h^2 \phi''(z_h) + 18 z_h^2 \phi'(z_h)^2 - 63 z_h \phi'(z_h) \,;
\nonumber \\ \cr && 
  B_2=-\sqrt{3}\bigg( 27\, z_h^2 \left( \phi'(z_h) \left( 2 z_h^2 \phi''(z_h) - 7 \right) + 2 z_h \phi'(z_h)^2 \right. \left. - 3 z_h \left( z_h \phi^{(3)}(z_h) + 3 \phi''(z_h) \right) \right)^2  \nonumber \ \cr &&  \quad \quad- 4 \left( -4 z_h^2 \phi''(z_h) + z_h^2 \phi'(z_h)^2 - 4 z_h \phi'(z_h) + 14 \right)^3 \bigg)^{1/2}\,.
\end{eqnarray}

\begin{figure}[!htb]
	\centering
	\includegraphics[scale=.4]{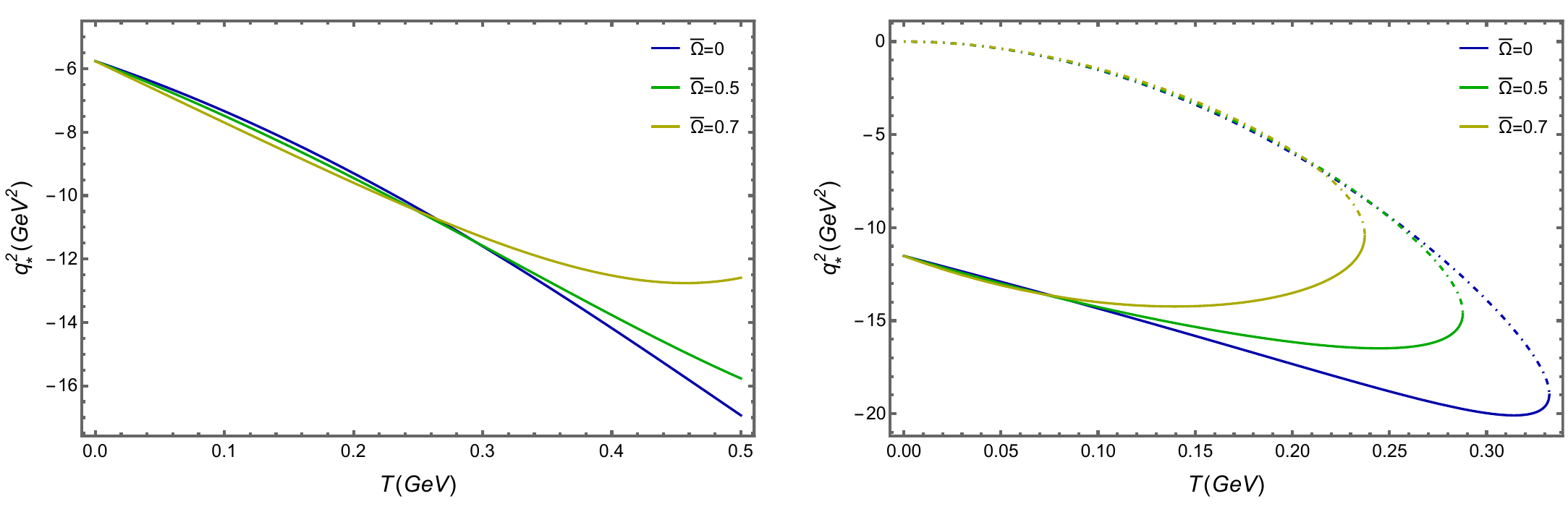}
	\caption{Pole-skipping values of the squared momentum as a function of temperature for the transverse case, computed for different values of $\bar{\Omega}$. The left panel shows the results for the first-order pole-skipping, while the right panel presents the results for the second order. In the right panel, the dot-dashed line represents the squared momentum $q_{*+}^2$, whereas the solid line corresponds to $q_{*-}^2$.}
    \label{PoleskippingsT}
\end{figure}

In Fig.~\ref{PoleskippingsT}, we show the pole-skipping values of the squared momentum \(q_{*}^{2}\) in the transverse channel as a function of temperature, computed for several values of the angular velocity. Unlike the longitudinal sector, where \(q_{*}\) remains real, the transverse momentum becomes complex. At leading order, \(q_{*}^{2}\) increases monotonically with temperature. On the other hand, increasing the angular velocity   decreases the absolute value of \(q_{*}^{2}\).

For the second-order, the behavior of the skipped poles exhibits a more intricate structure. Above a certain critical temperature, \(q_{*}^{2}\) acquires a non-zero imaginary part, corresponding to the collision between \(q_{*+}^{2}\) and \(q_{*-}^{2}\). Furthermore, this critical temperature decreases monotonically with increasing angular velocity.

 Finally, it is worth mentioning that the pole-skipping points in this case always occur at the Matsubara frequencies, $\omega_I = -2i\pi (n-1) T$ $(n=1,2,3,\dots)$, in agreement with similar results reported in the literature~\cite{Natsuume:2019sfp, Natsuume:2019xcy, Ahn:2019rnq, Ahn:2020baf, Wang:2022mcq, Pan:2024azf,Grozdanov:2019uhi,Jansen:2020hfd,Abbasi:2020xli}. 
 
\section{Phenomenological Implications}

Up to this point, we have discussed the physical properties of heavy quarkonia as modified by rotation in the case of nonvanishing momentum in the $x_3$ direction. To explore the experimental consequences of these effects, it is necessary to relate them to observables in heavy-ion collisions. In this section, we focus on the impact of rotation on the spin alignment of the $J/\psi$ meson.

In the framework of holography, the spectral function of vector mesons has recently been used to investigate the spin alignment of these particles in various media~\cite{Zhao:2024ipr,Sheng:2024kgg,Ahmed:2025bwi}. Following this approach, we estimate the spin alignment of the $J/\psi$ state in a rotating plasma. In contrast to Refs.~\cite{Zhao:2024ipr,Sheng:2024kgg,Ahmed:2025bwi}, we employ the spectral function expressed in terms of the electric field components in the longitudinal and transverse directions. This method significantly simplifies the calculation.

\subsection{Spin density matrix}

 The polarization properties of a vector meson can be characterized by a $3 \times 3$ spin density matrix, $\rho_{\lambda_1\lambda_2}$, where the indices $\lambda_1, \lambda_2 = 0, \pm 1$ denote the spin projections along the quantization axis. A widely used method to probe spin alignment is to analyze the meson's decay into a pair of leptons. For such two-body dilepton decays, the resulting angular distribution is given by \cite{Faccioli:2010kd}:
\begin{equation}\label{Wo}
    W(\theta) \propto \frac{1}{3+\lambda_\theta}\left(1 +\lambda_\theta \cos^2 \theta \right)
\end{equation}
where $\theta$ is the polar emission angle of the positively charged lepton with respect to the chosen quantization axis. For dilepton decays, the parameter $\lambda_\theta$ is related to the matrix element $\rho_{00}$ by
\begin{equation}\label{lambda}
   \lambda_\theta =\frac{1-3 \rho_{00}}{1+\rho_{00}}
\end{equation}
so that any deviation of $\lambda_\theta$ from zero signals a nontrivial spin alignment of the vector meson. 

The $S$-matrix element for the decay of a vector meson into a muon pair is
\begin{equation}
M_{fi} = \int d^4x \, d^4y \; \langle f, l \bar{l} | J_\mu^{l}(y) \, G_R(x-y) \, J_{\nu}^{l} (x)| i \rangle
\end{equation}
where $J_{\mu}(y)$ and $J^{l}_{\nu}$ are the hadronic and leptonic current respectively, and $G_{R}^{\mu \nu}$ is the retarded propagator of the vector meson at the vacuum: 
\begin{equation}
 G_R(p) =-\frac {\eta^{\mu\nu} + p^\mu p^\nu/p^2}{p^2 + m_V^2 + i m_V \Gamma}
\end{equation}
with $m_V$ the vacuum mass and $\Gamma$ is the width of the vector meson.

By summing over all possible final states and averaging over the initial state, one can express the total dimuon production rate as
\begin{equation}
n(x,p) = -\frac{2g_{M_{\mu^+\mu^-}}^2}{3(2\pi)^5} \left(1-\frac{2 m_\mu^2}{ p^2} \right) \sqrt{1+\frac{4m_{\mu}^2}{p^2}} p^2n_B(x, \omega)  \times
           \left( \eta^{\mu\nu} + \frac{p^\mu p^\nu}{p^2} \right) 
           G_{\mu\alpha}^{A}(p) \, \rho^{\alpha\beta}(x,p) \, G_{\beta\nu}^{R}(p)
\end{equation}
where $n_B(x, \omega)$ is the Bose-Einstein distribution,
\begin{equation}
n_B(x, \omega) = \frac{1}{e^{\omega / T(x)} - 1},
\end{equation}
with $T(x)$  the local temperature. Here, $m_\mu = 0.105\,\mathrm{GeV}$ is the muon mass and $g_{M_{\mu^+\mu^-}}$ denotes the coupling strength. The quantities $G^{R(A)}_{\mu\nu}$ correspond to the retarded (advanced) propagators in vacuum.

The spectral function in the medium is defined as
\begin{equation}
\rho^{\alpha\beta}(x,p) \equiv -\mathrm{Im}\, D^{\alpha\beta}(x,p)
\end{equation}
where  $D^{\alpha\beta}(x,p)$ is the retarded current-current correlator. The spectral function can be decomposed in terms of a complete set of polarization vectors, enabling the separation of contributions from each spin state:
\begin{equation}
\rho^{\mu\nu}(x, p) = \sum_{\lambda, \lambda' = 0, \pm 1} v^{\mu}(\lambda, p) v^{*\nu}(\lambda', p)\, \tilde{\rho}_{\lambda\lambda'}(x, p)
\end{equation}
where $v^\mu(\lambda,p)$ are the covariant polarization vectors, constructed as
\begin{equation}
v^\mu(\lambda,p) = \left(\frac{p \cdot \boldsymbol{\epsilon}_\lambda}{M},\, \boldsymbol{\epsilon}_\lambda + \frac{M}{\omega + M} \vec{p} \right)
\end{equation}
with $M \equiv \sqrt{\omega^2 - \vec{p}^2}$  the invariant mass of the vector meson. These polarization vectors satisfy the orthonormality and completeness relations:
\begin{align}
\eta_{\mu\nu} v^\mu(\lambda, p) v^{*\nu}(\lambda', p) &= \delta_{\lambda\lambda'} \\
\sum_\lambda v^\mu(\lambda, p) v^{*\nu}(\lambda, p) &= \eta^{\mu\nu} + \frac{p^\mu p^\nu}{p^2}
\end{align} 
where   $\boldsymbol{\epsilon}_\lambda$ represent the three possible spin orientations in the meson's rest frame, with $\boldsymbol{\epsilon}_0$ corresponds to the direction of spin quantization,  and $\boldsymbol{\epsilon}_{\pm 1}$ being perpendicular to $\boldsymbol{\epsilon}_0$. 

By projecting the spectral function onto the spin basis, one obtains:
\begin{equation}\label{specspin}
\tilde{\rho}_{\lambda\lambda'}(p) = \eta_{\mu\alpha} \eta_{\nu\beta} v^{*\alpha}(\lambda,p) v^\beta(\lambda',p) \rho^{\mu\nu}(p)
\end{equation}

The dilepton number yield for each spin state
\begin{equation}
n_\lambda(x,p) = - \frac{2g^2_{M_{\mu^+\mu^-}}}{3 (2\pi)^5} \left(1-\frac{2m_\mu^2}{p^2} \right) \times  \sqrt{1 - \frac{4m_\mu^2}{p^2}}  \frac{p^2 n_B(x, \omega)\tilde{\rho}_{\lambda\lambda}(x,p)}{(p^2 + m_V^2)^2 + m_V^2 \Gamma^2}
\end{equation}
The total dilepton number yield is given by summing over all spin states:
\begin{equation}
n =\sum_{\lambda=0, \pm 1} n_\lambda.
\end{equation}
The spin alignment of the vector meson corresponds to the probability of finding the system in a specific spin state, and for the produced dimuon, it is defined as
\begin{equation}\label{spinalign}
\rho_{\lambda\lambda'}(p) = - \frac{2g^2_{M_{\mu^+\mu^-}}}{3 (2\pi)^5} \left(1-\frac{2m_\mu^2}{p^2} \right) \times  \sqrt{1 - \frac{4m_\mu^2}{p^2}}  \frac{p^2 n_B(x, \omega)\tilde{\rho}_{\lambda\lambda}(x,p)}{(p^2 + m_V^2)^2 + m_V^2 \Gamma^2},
\end{equation}
with $N$ being the normalization factor that ensures the matrix is properly normalized ($N=\sum_{\lambda=0,\pm1}\rho_{\lambda\lambda}$), which guarantees that the sum of the diagonal of the spin state is unit. Note that for the restricted invariant mass near to the resonance mass $-p^2=m^2_{V}$, the spin alignment can be approximate to the ratio $\tilde{\rho}_{\lambda\lambda'}/\sum_{\lambda}\tilde{\rho}_{\lambda\lambda}$.

Therefore, using the spectral function derived from the holographic prescription, it is possible to study the spin alignment through the relation given in Eq.~(\ref{spinalign}).

\subsection{Spin Alignment in Holography}

In order to compute the spin alignment parameter $\lambda_\theta$ for the $J/\psi$ state, we consider the case where the ansatz for the wave vector takes the form
\begin{equation}\label{3momentum}
k^\mu = (-\omega, q_{\varphi}, q_{2}, q_{3})\,,
\end{equation}
where \( q_{\varphi} \) is the momentum along the \( \phi\) direction, \( q_{2} \) is the momentum along the \( x_2\) direction, \( q_{3} \) is the momentum along the \( x_3\) direction and \( \omega \) is the frequency. Consequently, the corresponding equations for the longitudinal (\ref{EqLRot2}) and transverse (\ref{EqTRot2}) components of the electric field take the following form:
\begin{eqnarray}\label{EqLRot2_final3d}
\partial_z^{2} \mathcal{E}_L &+& \left( -\frac{1}{z} + \frac{f'}{f} \frac{\gamma^2 (\omega- q_\varphi\bar{\Omega})^2}{\gamma^2 (\omega- q_\varphi\bar{\Omega})^2 - f(q_{2}^2 +q_{3}^2 + (q_\varphi-\omega \bar{\Omega})^2)} - \phi' \right) \partial_z \mathcal{E}_L \\ \cr   &+& \frac{\gamma^2 (\omega- q_\varphi\bar{\Omega})^2 - f(q_{2}^2 +q_{3}^2+ \gamma^2(q_\varphi-\omega \bar{\Omega})^2)}{f^2} \mathcal{E}_L = 0\,,
\end{eqnarray}
\begin{equation}\label{EqTRot2_final3d}
\partial_z^{2} \mathcal{E}_T + \left( -\frac{1}{z} + \frac{f'}{f} - \phi' \right) \partial_z \mathcal{E}_T + \frac{\gamma^2 (\omega- q_\varphi\bar{\Omega})^2 - f(q_2^2 +q_3^2+ \gamma^2(q_\varphi-\omega \bar{\Omega})^2)}{f^2} \mathcal{E}_T = 0\,.
\end{equation}
where $\mathcal{E}_{L/T}$ satisfies the infalling boundary condition (\ref{Infalling}). By following the approach described in Section~\ref{Retarded Green's Function}, we determine the spectral function in momentum space. The corresponding components of $\rho_{\mu \nu}$ are provided in Appendix~\ref{A2}.

Now, we choose the spin quantization axis along the momentum direction, corresponding to the helicity frame: $\boldsymbol{\epsilon}_0 = \vec{\mathbf{p}}/|\mathbf{p}|$. Substituting the spectral functions obtained from the holographic prescription into Eq.~(\ref{specspin}), the resulting spectral function in spin space takes the following form:
\begin{eqnarray}
\tilde{\rho}_{00} = Im  \Bigg[\lim_{z\rightarrow 0}\bigg(-\frac{e^{-\phi}\gamma^2\bar{\Omega}^2(q_2^2+q_3^{2})(q_2^2+q_3^2{+q_{\varphi}^2-\omega^2})}{2g_{5}^{2}(q_2^2+q_3^2+\gamma^2(q_{\varphi}-\omega\bar{\Omega})^2)z}\mathcal{E}_{T}(z)\mathcal{E}_{T}'(z) \nonumber \\ \cr + \frac{e^{-\phi}(q_2^2+q_3^2+q_{\varphi}(q_{\varphi}-\omega\bar{\Omega}))^2}{2g_{5}^{2}(q_2^2+q_3^2+q^2_{\phi})(q_2^2+q_3^2+\gamma^2(q_{\varphi}-\omega\bar{\Omega})^2)z}\mathcal{E}_{L}(z)\mathcal{E}_{L}'(z) \bigg) \Bigg] \,, \label{rho00} 
\end{eqnarray}

\begin{eqnarray}
\tilde{\rho}_{+1+1}=\tilde{\rho}_{-1-1} =Im   && \Bigg[\lim_{z\rightarrow 0}\bigg(\frac{ 
    2 (q_2^2 + q_3^2 + q_{\varphi}^2)^2 
    - 4 q_{\varphi} (q_2^2 + q_3^2 + q_{\varphi}^2) \, \omega \, \bar{\Omega}
    + (q_2^2 + q_3^2 + 2 q_{\varphi}^2) \omega^2 \bar{\Omega}^2
}{2 (q_2^2 + q_3^2 + q_{\varphi}^2) \left(   + q_2^2  + q_3^2 +\gamma^2(q_{\varphi} - \omega \bar{\Omega})^2\right)} \nonumber \\  \cr && -\frac{ 
      (q_2^2 + q_3^2) (q_2^2 + q_3^2 + q_{\varphi}^2) \, \bar{\Omega}^2
    }{2 (q_2^2 + q_3^2 + q_{\varphi}^2) \left(   + q_2^2  + q_3^2 +\gamma^2(q_{\varphi} - \omega \bar{\Omega})^2\right)}\bigg) \frac{e^{-\phi}}{2g_{5}^{2}z}\mathcal{E}_{T}(z)\mathcal{E}_{T}'(z) \nonumber \\  \cr &&- \lim_{z\rightarrow 0}\bigg( \frac{e^{-\phi}\gamma^2\bar{\Omega}^2(q_2^2+q_3^{2})(q_2^2+q_3^2{+q_{\varphi}^2-\omega^2})}{2g_{5}^{2}(q_2^2+q_3^2+\gamma^2(q_{\varphi}-\omega\bar{\Omega})^2)z}\mathcal{E}_{L}(z)\mathcal{E}_{L}'(z) \bigg) \Bigg]
    \label{rho11}
\end{eqnarray} 

\begin{figure}[!htb]
	\centering
	\includegraphics[scale=.5]{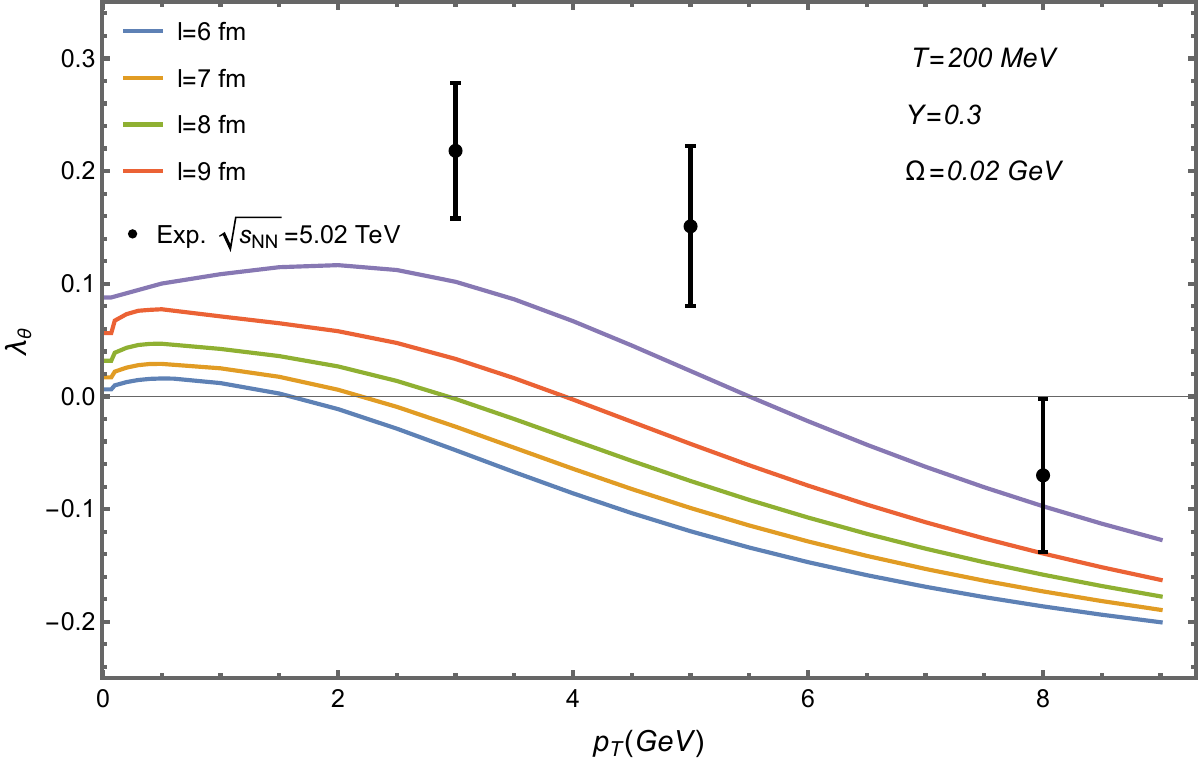}
	\caption{The spin alignment parameter $\lambda_\theta$ for charmonium is presented for different values of $l$ at $T = 200$ MeV, $Y = 0.3$, and $\Omega = 0.02$ GeV.}
    \label{SpinAlignmentx}
\end{figure}

Finally, we can compute the spin alignment by employing the standard parameterization of the meson momentum used in heavy-ion collisions, namely in terms of the transverse momentum $p_T$, azimuthal angle $\alpha$, and rapidity $Y$:
\begin{equation}\label{heavyion}
k^\mu = \left( \sqrt{M^2+p_T^2} \cosh Y,\, p_T \sin\alpha,\, p_T \cos\alpha,\, \sqrt{M^2+p_T^2}\sinh Y \right).
\end{equation}
For comparison with experimental values, we integrate over the azimuthal angle. The spin alignment parameter $\lambda_\theta$ is then computed using Eq. (\ref{spinalign}), evaluated near the resonance mass, together with the relation in Eq. (\ref{lambda}). The results for $\lambda_\theta$ as a function of $p_T$ at $Y = 0.3$, for different values of $l$, are presented in Fig. \ref{SpinAlignmentx}, together with experimental data from Refs.\cite{ALICE:2020iev,ALICE:2022dyy}.

We observe qualitative agreement between our results and experimental measurements as the radius $l$ increases. This prediction could be tested in future experiments to investigate how spin alignment is affected by variations in angular velocity or radius $l$ in non-central heavy-ion collisions. Another point to note is that our results are more consistent with experimental data at higher values of $p_T$. This behavior is expected, since the available experimental data for $\lambda_\theta$ are limited to the forward rapidity region ($2.5 \leq Y \leq 4$), whereas our results correspond to the mid-rapidity region. Future measurements at mid-rapidity would allow for a more direct comparison with theoretical predictions. It would also be interesting to include the effects of the magnetic field, as explored in Ref.~\cite{Zhao:2024ipr}.

\section{Conclusion and Discussion}

In this work, we have investigated the real and imaginary parts of the quasinormal modes frequencies associated with charmonium states in a holographic bottom-up model. We studied their temperature dependence for quasi-particles moving relative to the rotating medium. In addition, we calculated the diffusion constant in the rotating plasma and analyzed the pole-skipping phenomenon, which emerges in the hydrodynamic modes of the holographic charmonium model. Lastly, we have calculated the spin alignment parameter $\lambda_{\theta}$ to investigate the effect of rotation in non-central collisions.

The results obtained from the quasinormal modes are consistent with the behavior observed in the spectral function, as we now discuss. For $J/\psi$ moving relative to the rotating medium, the real part of the quasinormal  frequencies increases with momentum, for both longitudinal and transverse motion relative to the wave vector. This is translated in terms of the spectral function into the increase in the value of the frequency where the peaks are located. Moreover, when the momentum is fixed and the angular velocity approaches 1, the real part of the frequency increases rapidly in the longitudinal direction, whereas it decreases in the transverse direction. For the imaginary part of the frequency, the effect of rotation is similar in both directions. The differences between the longitudinal and transverse cases emerge as momentum increases: in the longitudinal case, the imaginary part decreases with increasing momentum, while in the transverse case, it increases. This behaviour is consistently reproduced in the spectral function as an increase in the height of the peaks. For the excited states, an investigation of the spectral function and quasinormal modes at very low temperatures and angular velocities is required, as these states dissociate rapidly. In the temperature range considered in this work, the excited states are already melted. Notice also that our results indicate the size of the QGP may play a significant role in the dissociation processes, as increasing the radius of the hyper-cylinder leads to an enhancement of dissociation.

Throughout this paper, we have investigated the effects of rotation on $J/\psi$ state by analyzing the spectral function and quasinormal modes. However, the relevant experimental observable for charmonium suppression is the nuclear modification factor, $R_{AA}$. Although it is possible to calculate this observable using the spectral function by extracting the thermal width\footnote{ It is important to note that the width in our case can be determined using the quasinormal modes.}, as done in Refs.~\cite{Blaizot:2021xqa,Thakur:2021vbo}, a meaningful comparison with experiment would require averaging over polarization states, since our calculations have separated transverse and longitudinal polarizations, a distinction that is not experimentally accessible at present. Based on the results presented here, we speculate that $R_{AA}$ should decrease as a function of angular velocity or radius. Similarly, we expect $R_{AA}$ to decrease with increasing momentum, as observed in Ref.~\cite{Hohler:2013vca}.

To more clearly reveal the effect of rotation on an observable, we calculate the spin alignment parameter $\lambda_\theta$ using the spectral function derived from holography. We evaluated $\lambda_\theta$ for different values of $l$, and our predictions indicate that increasing $l$ leads to qualitative agreement with the experimental data~\cite{ALICE:2020iev,ALICE:2022dyy}. Nevertheless, the available data for $\lambda_\theta$ are limited to the forward rapidity region, whereas our analysis is focused on the mid-rapidity region. This prediction could be tested in future experiments to investigate how spin alignment is affected by variations in angular velocity or radius $l$ in non-central heavy-ion collisions. It would be interesting to analyze the combined effects of angular velocity and magnetic field on $\lambda_\theta$, given that strong magnetic fields are also present in non-central collisions. We leave this investigation for future work.

In addition, the charge diffusion coefficient derived  exhibits a non-trivial dependence on temperature, in contrast to the trivial temperature dependence found in the conformal AdS/CFT case. However, the effect of rotation is simpler: It rescales the diffusion coefficient by a factor of $\gamma$, consistent with the results obtained in the $AdS_4$ case~\cite{Morgan:2013dv}.

Another phenomenon analyzed in this paper is the  pole-skipping of the holographic Green's functions. In the longitudinal sector, pole-skipping points lie on the hydrodynamic modes of the model, where the momentum is real. In contrast, the skipped poles in the transverse sector occur at complex momentum values. Moreover, the effect of rotation manifests solely through the momentum values. Due to the relation between the temperature and the position of the horizon, the frequency does not explicitly depend on the angular velocity, but rather only on the temperature. Notice that the pole-skipping points do not exhibit any relationship with the charmonium quasinormal modes spectrum. Additionally, these points have no connection with the physical region, even in holographic models featuring a mass gap, which is characteristic of more realistic scenarios.

Furthermore, expressions (\ref{poleLong}) and (\ref{firstorder}) provide the identification of pole-skipping points for other dilaton profiles in the soft wall model. If the dilaton function is chosen to take negative values, pole-skipping points may no longer arise in the hydrodynamic modes of the model, and the corresponding points in the transverse sector may instead occur at real momentum values. Nevertheless, the frequency remains unaffected and is independent of the specific dilaton profile.

Finally, we remark that the authors of Ref. \cite{Braga:2023fac} investigated the quasinormal modes of bottomonium in a rotating plasma; however, their analysis was restricted to the case $q=0$. Similarly, Ref. \cite{Zhao:2023pne} focused on the spectral function of charmonium, also restricted to $q=0$. Moreover, neither of these studies considered the equations of motion in the longitudinal or transverse directions relative to the wave vector.

\paragraph*{\textbf{Acknowledgments}: L.F.F is supported by ANID Fondecyt postdoctoral grant folio No. 3220304.}

\appendix

\section{Derivation of the Pole-Skipping}
\label{A1}

The method developed in \cite{Blake:2019otz} involves analyzing the near-horizon behavior of perturbations using Eddington-Finkelstein (EF) coordinates. Then, in order to determine the pole-skipping points for the vector field in the soft-wall model, we formulate the equations of motion separately for the transverse and longitudinal components in EF coordinates.

The equation for the longitudinal component is given by:
\begin{eqnarray}\label{EFL}
    && \hfill  E_{L}''+E_{L}''\Bigg(\frac{f'}{f} 
    \frac{\gamma^2 \omega^2 }{ \left(\gamma^2 \omega^2 - f \left(q^2+\gamma^2 \omega^2 \bar{\Omega}^2  \right) \right)} 
    + \frac{2 i \gamma \omega}{f} - \phi' - \frac{1}{z} 
    \Bigg) +  E_{L}'\Bigg(-\frac{q^2+\gamma^2\omega^2\bar{\Omega}^2}{f}  \nonumber \\
    && - \frac{i \gamma \omega }{f}\left(  \frac{1}{z} +\phi'-\frac{f'(q^2+\gamma^2\omega^2\bar{\Omega}^2)}{\gamma^2\omega^2-f(q^2+\gamma^2 \omega^2 \bar{\Omega}^2 )}\right)\Bigg) =0
\end{eqnarray}
Similarly, for the transverse component, we have:
\begin{eqnarray}\label{EFT}
    &&E_T''+  E_T' \left( -\frac{1}{z} + \frac{2 i \gamma \omega}{f} 
    + \frac{f'}{f} - \phi' \right)  + E_T\Bigg( -\frac{q^2+\gamma^2\omega^2\bar{\Omega}^2}{f}  - \frac{i \gamma \omega }{f}\left(  \frac{1}{z} +\phi'\right) \Bigg)=0
\end{eqnarray}
Here, the prime denotes differentiation with respect to the coordinate $z$. For simplicity, we omit the index $\mu$ in the electric field components.

Next, we rearrange these equations into a more convenient form to facilitate the analysis. The longitudinal component equation \eqref{EFL} can be rewritten as:
\begin{eqnarray}\label{EFL2x}
E_L''+ \Xi_{L}(z)  E_L'  + \beta_{L}(z) E_L=0
\end{eqnarray}
and the transverse component equation \eqref{EFT} takes the form:
\begin{eqnarray}\label{EFT2x}
    &&E_T''+ \Xi_{T}(z)  E_T'  + \beta_{T}(z) E_T=0
\end{eqnarray}
where  $\Xi_{L/T}(z)$ and $\beta_{L/T}(z)$ are coefficients. The near-horizon solutions of equations \eqref{EFL2x} and \eqref{EFT2x} take the following form:
%Then, we expand the solution around the horizon as follows:
\begin{equation}
    E_{L/T}=(z-z_h)^{\alpha}\sum_{m=0}^{\infty} E_{(m)L/T}(z-z_h)^{m}
\end{equation}
where $E_{(m)L/T}$ are expansion coefficients determined recursively and $\alpha$ is a characteristic exponent that determines the leading-order behavior of the solution.

The two coefficients $\Xi_{L/T}(z)$ are also expanded near the horizon and must satisfy the following expansions:
\begin{eqnarray}
&&\Xi_{L/T}(z)=\frac{\Xi_0}{z-z_h}+\Xi_1+\Xi_2(z-z_h)+O(z-z_h)^2 \cr
&&\beta_{L/T}(z)=\frac{\beta_0}{z-z_h}+\beta_1+\beta_2(z-z_h)+O(z-z_h)^2 
\end{eqnarray}
where $\Xi_i$ and $\beta_{i}$ are coefficients that depend on  $\omega$, $q$ and $\bar{\Omega}$.

Plugging these expressions into equations (\ref{EFL2x}) and (\ref{EFT2x}), we obtain the corresponding series expansions for each equation as follows:
\begin{equation}\label{Expansion2}
    \mathfrak{g_0}(z-z_h)^{\lambda-2}+\mathfrak{g_1}(z-z_h)^{\lambda-1}+\mathfrak{g_2}(z-z_h)^{\lambda}+\dots=0
\end{equation}
where
\begin{eqnarray}\label{Eqs}
&&\mathfrak{g_0}=\lambda(\lambda-1+\Xi_0)E_0=0\,,\cr && \mathfrak{g_1}=(\lambda+1)(\lambda+\Xi_0)E_1+(\lambda\Xi_1+\beta_0)E_0=0\,,\cr && \mathfrak{g_2}=(\lambda+2)(\lambda+1+\Xi_0)E_2+((\lambda+1)\Xi_1+\beta_0)E_1+(\lambda\Xi_2+\beta_1)=0\,, \cr && \cdots
\end{eqnarray}
We then proceed to solve for the coefficients $F_i$, $i=1,2\cdots$ order by order. At zeroth order, we find that $\alpha$ admits two possible values:
\begin{equation}
    \lambda_1=0\,,\,\,\,\,\,\,\,\, \lambda_2=\frac{i\omega}{\pi T}
\end{equation}
The first solution, $\lambda_1=0$, corresponds to the ingoing solution, which is physically relevant for causal propagation at the horizon. The second solution, $\lambda_2=\frac{i\omega}{\pi T}$ corresponds to the outgoing solution, which is typically discarded in the standard prescription for computing retarded Green’s functions in holography.

Substituting the ingoing solution into the first-order equation, we obtain:
\begin{eqnarray}
\Xi_0E_1+\beta_0E_0=0\,.    
\end{eqnarray}
Since $E_1$ can generally be expressed in terms of $E_0$, the higher-order coefficients $E_j$  can be determined iteratively, allowing for a systematic construction of the full solution near the horizon.

As demonstrated in \cite{Blake:2019otz,Pan:2024azf}, when the coefficients $\Xi_0$ and $\beta_0$ vanish, the ingoing solution becomes non-unique. As a result, pole-skipping occurs. Consequently, the pole-skipping points can be identified by imposing the following condition:
\begin{equation} \Xi_0=\beta_0=0 \end{equation}
This implies that the specific values of frequency and momentum at which these coefficients vanish correspond to pole-skipping. In this case, they determine the leading pole-skipping point.

More pole-skipping points can be identified by incorporating higher-order terms in the equations, thus extending the analysis beyond the leading-order case:
\begin{equation} \Xi_n E_n + \beta_n E_0 = 0. \end{equation}
Consequently, the pole-skipping points $(\omega_*,q_*)$ can be determined by finding the frequency and momentum at which the coefficients satisfy the condition:
\begin{equation} E_0 = E_n = 0. \end{equation}
This approach systematically reveals additional pole-skipping points, further enriching the understanding of the Green’s function structure.

\section{Components of the Spectral function}
\label{A2}

Using the same procedure outlined in Section~\ref{Retarded Green's Function}, we derive the spectral function for the case of nonzero momentum, as given in Eq.~(\ref{3momentum}). The components are presented below:
\begin{eqnarray}\label{rhott}
\rho_{tt}=-Im \lim_{z\rightarrow0}\Bigg[\frac{e^{-\phi}f}{2g_{5}^{2}z}\Bigg(&&\frac{(q_2^2+q_3^2+q_{\varphi}^2-q_{\varphi}\omega\bar{\Omega})^2\gamma^2}{\left( \omega^2-q_2^2-q_3^2-q_{\varphi}^2\right)(q_2^2+q_3^2+\gamma^2(q_\varphi-\omega\bar{\Omega})^2)}\mathcal{E}_{L}(z)\mathcal{E}_{L}'(z) \nonumber \\ \cr &&+\frac{(q_2^2+q_3^2)\gamma^2\bar{\Omega}^2}{(q_2^2+q_3^2+\gamma^2(q_\varphi-\omega\bar{\Omega})^2)}\mathcal{E}_{T}(z)\mathcal{E}_{T}'(z)\Bigg)\Bigg]
\end{eqnarray}

\begin{eqnarray}\label{rhotx2}
\rho_{tx_2}=\rho_{x_2t}=-Im \lim_{z\rightarrow0}\Bigg[\frac{e^{-\phi}f}{2g_{5}^{2}z}\Bigg(&&\frac{
q_2 \, \gamma^2 \left( -\omega + q_{\varphi} \, \bar{\Omega} \right) 
\left( q_2^2 + q_3^2 + q_{\varphi} \left( q_{\varphi} - \omega \, \bar{\Omega} \right) \right)
}{
\left( q_2^2 + q_3^2 + q_{\varphi}^2 - \omega^2 \right)
\left( q_2^2 + q_3^2 + \gamma^2 \left( q_{\varphi} - \omega \, \bar{\Omega} \right)^2 \right)
}\mathcal{E}_{L}(z)\mathcal{E}_{L}'(z) \nonumber \\ \cr &&+\frac{q_2 \, \gamma^2 \, \bar{\Omega} \left( -q_{\varphi} + \omega \, \bar{\Omega} \right)}{q_2^2 + q_3^2 + \gamma^2 \left( q_{\varphi} - \omega \, \bar{\Omega} \right)^2}
\mathcal{E}_{T}(z)\mathcal{E}_{T}'(z)\Bigg)\Bigg]
\end{eqnarray}

\begin{eqnarray}\label{rhotx3}
\rho_{tx_3}=\rho_{x_3 t}=-Im \lim_{z\rightarrow0}\Bigg[\frac{e^{-\phi}f}{2g_{5}^{2}z}\Bigg(&&\frac{
q_3 \, \gamma^2 \left( -\omega + q_{\varphi} \, \bar{\Omega} \right) 
\left( q_2^2 + q_3^2 + q_{\varphi} \left( q_{\varphi} - \omega \, \bar{\Omega} \right) \right)
}{
\left( q_2^2 + q_3^2 + q_{\varphi}^2 - \omega^2 \right)
\left( q_2^2 + q_3^2 + \gamma^2 \left( q_{\varphi} - \omega \, \bar{\Omega} \right)^2 \right)
}\mathcal{E}_{L}(z)\mathcal{E}_{L}'(z) \nonumber \\ \cr &&+\frac{q_3 \, \gamma^2 \, \bar{\Omega} \left( -q_{\varphi} + \omega \, \bar{\Omega} \right)}{q_2^2 + q_3^2 + \gamma^2 \left( q_{\varphi} - \omega \, \bar{\Omega} \right)^2}
\mathcal{E}_{T}(z)\mathcal{E}_{T}'(z)\Bigg)\Bigg]
\end{eqnarray}

\begin{eqnarray}\label{rhotphi}
\rho_{t\varphi}=\rho_{\varphi t}=-Im \lim_{z\rightarrow0}\Bigg[\frac{e^{-\phi}f}{2g_{5}^{2}z}\Bigg(&&- \frac{
\gamma^2 \left( q_{\varphi} \, \omega + \left( q_2^2 + q_3^2 - \omega^2 \right) \bar{\Omega} \right)
\left( q_2^2 + q_3^2 + q_{\varphi} \left( q_{\varphi} - \omega \, \bar{\Omega} \right) \right)
}{
\left( q_2^2 + q_3^2 + q_{\varphi}^2 - \omega^2 \right)
\left( q_2^2 + q_3^2 + \gamma^2 \left( q_{\varphi} - \omega \, \bar{\Omega} \right)^2 \right)
}\mathcal{E}_{L}(z)\mathcal{E}_{L}'(z) \nonumber \\ \cr &&+\frac{(q_2^2 + q_3^2) \, \gamma^2 \, \bar{\Omega}}{q_2^2 + q_3^2 + \gamma^2 \left( q_{\varphi} - \omega \, \bar{\Omega} \right)^2}
\mathcal{E}_{T}(z)\mathcal{E}_{T}'(z)\Bigg)\Bigg]
\end{eqnarray}

\begin{eqnarray}\label{rhox2x2}
\rho_{x_2x_2}=-Im \lim_{z\rightarrow0}\Bigg[\frac{e^{-\phi}f}{2g_{5}^{2}z}\Bigg(&&- \frac{
q_2^2 \, \gamma^2 \left( \omega - q_{\varphi} \, \bar{\Omega} \right)^2
}{
\left( q_2^2 + q_3^2 + q_{\varphi}^2 - \omega^2 \right)
\left( q_2^2 + q_3^2 + \gamma^2 \left( q_{\varphi} - \omega \, \bar{\Omega} \right)^2 \right)
}\mathcal{E}_{L}(z)\mathcal{E}_{L}'(z) \nonumber \\ \cr &&+\frac{
q_3^2 + \gamma^2 \left( q_{\varphi} - \omega \, \bar{\Omega} \right)^2
}{
q_2^2 + q_3^2 + \gamma^2 \left( q_{\varphi} - \omega \, \bar{\Omega} \right)^2
}\mathcal{E}_{T}(z)\mathcal{E}_{T}'(z)\Bigg)\Bigg]
\end{eqnarray}

\begin{eqnarray}\label{rhox2phi}
\rho_{x_2\varphi}=\rho_{\varphi x_2}=-Im \lim_{z\rightarrow0}\Bigg[\frac{e^{-\phi}f}{2g_{5}^{2}z}\Bigg(&&\frac{
q_2 \, \gamma^2 \left( -\omega + q_{\varphi} \, \bar{\Omega} \right)
\left( q_{\varphi} \, \omega + \left( q_2^2 + q_3^2 - \omega^2 \right) \bar{\Omega} \right)
}{
\left( q_2^2 + q_3^2 + q_{\varphi}^2 - \omega^2 \right)
\left( q_2^2 + q_3^2 + \gamma^2 \left( q_{\varphi} - \omega \, \bar{\Omega} \right)^2 \right)
}\mathcal{E}_{L}(z)\mathcal{E}_{L}'(z) \nonumber \\ \cr &&+\frac{
q_2 \, \gamma^2 \left( -q_{\varphi} + \omega \, \bar{\Omega} \right)
}{
q_2^2 + q_3^2 + \gamma^2 \left( q_{\varphi} - \omega \, \bar{\Omega} \right)^2
}\mathcal{E}_{T}(z)\mathcal{E}_{T}'(z)\Bigg)\Bigg]
\end{eqnarray}

\begin{eqnarray}\label{rhox2x3}
\rho_{x_2 x_3}=\rho_{x_3 x_2}=-Im \lim_{z\rightarrow0}\Bigg[\frac{e^{-\phi}f}{2g_{5}^{2}z}\Bigg(&&- \frac{
q_2 \, q_3 \, \gamma^2 \left( \omega - q_{\varphi} \, \bar{\Omega} \right)^2
}{
\left( q_2^2 + q_3^2 + q_{\varphi}^2 - \omega^2 \right)
\left( q_2^2 + q_3^2 + \gamma^2 \left( q_{\varphi} - \omega \, \bar{\Omega} \right)^2 \right)
}\mathcal{E}_{L}(z)\mathcal{E}_{L}'(z) \nonumber \\ \cr &&- \frac{
q_2 \, q_3
}{
q_2^2 + q_3^2 + \gamma^2 \left( q_{\varphi} - \omega \, \bar{\Omega} \right)^2
}\mathcal{E}_{T}(z)\mathcal{E}_{T}'(z)\Bigg)\Bigg]
\end{eqnarray}

\begin{eqnarray}\label{rhox3x3}
\rho_{x_3 x_3}=-Im \lim_{z\rightarrow0}\Bigg[\frac{e^{-\phi}f}{2g_{5}^{2}z}\Bigg(&&-\frac{q_3^2 \, \gamma^2 \left( \omega - q_{\varphi} \, \bar{\Omega} \right)^2}{\left( q_2^2 + q_3^2 + q_{\varphi}^2 - \omega^2 \right)
\left( q_2^2 + q_3^2 + \gamma^2 \left( q_{\varphi} - \omega \, \bar{\Omega} \right)^2 \right)
}\mathcal{E}_{L}(z)\mathcal{E}_{L}'(z) \nonumber \\ \cr && \frac{
q_2^2 + \gamma^2 \left( q_{\varphi} - \omega \, \bar{\Omega} \right)^2
}{
q_2^2 + q_3^2 + \gamma^2 \left( q_{\varphi} - \omega \, \bar{\Omega} \right)^2
}\mathcal{E}_{T}(z)\mathcal{E}_{T}'(z)\Bigg)\Bigg]
\end{eqnarray}

\begin{eqnarray}\label{rhox3phi}
\rho_{x_3 \varphi}=\rho_{\varphi x_3}=-Im \lim_{z\rightarrow0}\Bigg[\frac{e^{-\phi}f}{2g_{5}^{2}z}\Bigg(&&\frac{
q_3 \, \gamma^2 \left( -\omega + q_{\varphi} \, \bar{\Omega} \right)
\left( q_{\varphi} \, \omega + \left( q_2^2 + q_3^2 - \omega^2 \right) \bar{\Omega} \right)
}{
\left( q_2^2 + q_3^2 + q_{\varphi}^2 - \omega^2 \right)
\left( q_2^2 + q_3^2 + \gamma^2 \left( q_{\varphi} - \omega \, \bar{\Omega} \right)^2 \right)
}\mathcal{E}_{L}(z)\mathcal{E}_{L}'(z) \nonumber \\ \cr && \frac{
q_3 \, \gamma^2 \left( -q_{\varphi} + \omega \, \bar{\Omega} \right)
}{
q_2^2 + q_3^2 + \gamma^2 \left( q_{\varphi} - \omega \, \bar{\Omega} \right)^2
}\mathcal{E}_{T}(z)\mathcal{E}_{T}'(z)\Bigg)\Bigg]
\end{eqnarray}

\begin{eqnarray}\label{rhophiphi}
\rho_{\varphi \varphi}=-Im \lim_{z\rightarrow0}\Bigg[\frac{e^{-\phi}f}{2g_{5}^{2}z}\Bigg(&&- \frac{
\gamma^2 \left( q_{\varphi} \, \omega + \left( q_2^2 + q_3^2 - \omega^2 \right) \bar{\Omega} \right)^2
}{
\left( q_2^2 + q_3^2 + q_{\varphi}^2 - \omega^2 \right)
\left( q_2^2 + q_3^2 + \gamma^2 \left( q_{\varphi} - \omega \, \bar{\Omega} \right)^2 \right)
}\mathcal{E}_{L}(z)\mathcal{E}_{L}'(z) \nonumber \\ \cr && \frac{
(q_2^2 + q_3^2) \, \gamma^2
}{
q_2^2 + q_3^2 + \gamma^2 \left( q_{\varphi} - \omega \, \bar{\Omega} \right)^2
}\mathcal{E}_{T}(z)\mathcal{E}_{T}'(z)\Bigg)\Bigg]
\end{eqnarray}

\bibliographystyle{apsrev4-2}
\bibliography{refs}

\end{document}